\documentclass[twocolumn,showpacs,superscriptaddress,prb]{revtex4-1}
\usepackage{amsfonts}
\usepackage{amsmath}
\usepackage{amssymb}
\usepackage{graphicx}

\begin{document}

\title[Magnetooscillations and vanishing resistance states]{Microwave-resonance-induced magnetooscillations and vanishing resistance
states in multisubband two-dimensional electron systems}
\author{Yu.P. Monarkha}
\affiliation{Institute for Low Temperature Physics and Engineering, 47 Lenin Avenue, 61103
Kharkov, Ukraine}

\begin{abstract}
The dc magnetoconductivity of the multisubband two-dimensional electron system
formed on the liquid helium surface in the presence of resonant microwave
irradiation is described, and a new mechanism of the negative linear response
conductivity is studied using the self-consistent Born approximation. Two
kinds of scatterers (vapor atoms and capillary wave quanta) are considered.
Besides a conductivity modulation expected near the points, where the
excitation frequency for inter-subband transitions is commensurate with the
cyclotron frequency, a sign-changing correction to the linear
conductivity is shown to appear for usual quasi-elastic inter-subband
scattering, if the collision broadening of Landau levels is much smaller than
thermal energy. The decay heating of the electron system near the
commensurability points leads to magnetooscillations of electron temperature,
which are shown to increase the importance of the sign-changing correction.
The line-shape of magnetoconductivity oscillations calculated for wide ranges
of temperature and magnetic field is in a good accordance with experimental observations.
\end{abstract}









\maketitle



\section{Introduction}

The discovery of novel microwave-induced oscillations of magnetoresistivity~\cite{ZudSim-01}
as a function of the magnetic field $B$ and so-called
zero-resistance states
(ZRS)~\cite{ManSmet-02,ZudDu-03} has sparked a large interest in quantum
magnetotransport of two-dimensional (2D) electron systems exposed to microwave (MW)
radiation. The 1/B-periodic oscillations were observed for quite arbitrary
MW\ frequencies $\omega$ larger than the cyclotron frequency $\omega_{c}$. The
period of these oscillations is governed by the ratio $\omega/\omega_{c}$. ZRS
appear in ultrahigh-mobility GaAs/AlGaAs heterostructures as a result of
evolution of the minima of the oscillations with an increase in radiation power.

Recently, MW-induced magnetooscillations and vanishing of the magnetoconductance
$\sigma_{xx}$ were observed in the nondegenerate multisubband 2D electron
system formed on the liquid helium surface~\cite{KonKon-09,KonKon-10}. These
oscillations have many striking similarities with those observed in
semiconductor systems: they are 1/B-periodic, governed by the ratio
$\omega/\omega_{c}$, and their minima eventually evolve in zero
magnetoconductance states nearly at the same values of
$\omega/\omega_{c}$. The important distinction of
these new oscillations is that they are observed only for a MW frequency fixed to the
resonance condition for excitation of the second surface subband: $\hbar
\omega=\Delta_{2,1}$ (here $\Delta_{l,l^{\prime}}=\Delta_{l}-\Delta
_{l^{\prime}}$, and $\Delta_{l}$ describes the energy spectrum of surface
electron states, $l=1,2,...$).

The ZRS observed in semiconductor systems are shown~\cite{And-03} to be
understood as a direct consequence of the negative photoconductivity $\sigma
_{xx}<0$ which can appear with an increase in the amplitude of conductivity
oscillations. Regarding the microscopic origin of the oscillations, the most
frequently studied mechanism is based on photon-induced impurity scattering
within the ground subband, when an electron simultaneously is scattered off
impurities and absorb or emit microwave quanta~\cite{Ryz-69,DurSac-03}. This kind of
scattering is accompanied by an electron displacement along the applied
dc-electric field whose sign depends on the sign of $\omega_{c}\left(
n-n^{\prime}\right)  +\omega$ (here $n=0,1,...$). Therefore, sometimes this
mechanism is termed the "displacement"
mechanism. A different mechanism, called the "inelastic"
mechanism~\cite{Dmi-07}, explains conductivity oscillations as a result of
oscillatory changes of the isotropic part of the in-plane electron
distribution function.

Both microscopic mechanisms of the negative conductivity prosed for
semiconductor systems cannot be applied for explanation of
similar effects observed in the system of surface electrons (SEs) on liquid
helium, because the MW frequency considered in these theories has no relation
to inter-subband excitation frequencies $\Delta_{l,l^{\prime}}/\hbar
\equiv\omega_{l,l^{\prime}}$. Recently, a new mechanism of negative momentum
dissipation relevant to experiments with SEs on liquid helium
was briefly reported~\cite{Mon-11}. It cannot be attributed to "displacement"
or "inelastic" mechanisms. In this theory, the origin of
magnetooscillations and negative dissipation is an additional filling of the
second surface subband induced by MW irradiation under the resonance condition
($\hbar\omega=\Delta_{2,1}$), which triggers quasi-elastic inter-subband
electron scattering. The ordinary inter-subband scattering, which does not
involve photon quanta, is accompanied by electron displacements whose sign
depends on the sign of $\omega_{c}\left(  n-n^{\prime}\right)  +\omega_{2,1}$.
Usually, this scattering does not lead to any negative contribution to
$\sigma_{xx}$
. A correction to $\sigma_{xx}$ proportional to $\omega_{c}\left(  n-n^{\prime}\right)
+\omega_{2,1}$ was shown to appear only if $N_{2}>N_{1}e^{-\Delta_{2,1}/T_{e}%
}$, where $N_{l}$ is the number of electrons at the corresponding subband, and
$T_{e}$ is the electron temperature. It is important that this correction is
also proportional to a large parameter equal to the ratio of $T_{e}$
to the collision broadening of Landau levels.

In this work, we perform a systematic theoretical study of negative
dissipation phenomena in a multisubband 2D electron system caused by
non-equilibrium filling of excited subbands. The magnetotransport
theory~\cite{Mon-11} is generalized in order to include electron
scattering by capillary wave quanta (ripplons) which limits SE mobility in
experiments~\cite{KonKon-10}, where vanishing magnetoconductivity $\sigma_{xx}$ is
observed. In order to understand the importance of MW heating at the vicinity of
commensurability points, electron energy relaxation is analyzed.
A sign-changing correction to the energy relaxation rate
similar to the sign-changing correction to the momentum relaxation rate is found
for non-equilibrium filling of excited subbands.

\section{General definitions}

Consider a multisubband 2D electron system under magnetic field applied
perpendicular. The electron energy spectrum is described by
$\Delta_{l}+\varepsilon_{n}$, where $\varepsilon_{n}=\hbar\omega_{c}\left(
n+1\right)  $ represents Landau levels ($n=0,1,2...$). For SEs on liquid helium
(for review see Ref.~\onlinecite{MonKon-04}) under
a weak holding electric field ($E_{\bot}\rightarrow0$), $\Delta_{l}\simeq-$
$E_{R}/l^{2}$, where $E_{R}$ is the effective Rydberg energy of SE states,
\begin{equation}
E_{R}=\frac{\hbar^{2}}{2m_{e}a_{B}^{2}},\text{ \ }a_{B}=\frac{\hbar^{2}}%
{m_{e}\Lambda},\text{ \ }\Lambda=\frac{e^{2}\left(  \epsilon-1\right)
}{4\left(  \epsilon+1\right)  }, \label{e1}%
\end{equation}
$a_{B}$ is the effective Bohr radius, $m_{e}$ is the electron mass,
and $\epsilon$ is the dielectric
constant of liquid helium. The excitation energy $\Delta_{2,1}$ is about
$6\,\mathrm{K}$ (liquid $^{4}\mathrm{He}$) or $3.2\,\mathrm{K}$ (liquid
$^{3}\mathrm{He}$). It increases with the holding electric field $E_{\bot}$,
which allows also to tune $\Delta_{2,1}$ in resonance with the MW frequency.

Under typical experimental conditions, the electron-electron collision rate
$\nu_{e-e}$ of SEs is much higher than the energy and momentum relaxation
rates. Therefore, the electron distribution as a function of the
in-plane energy $\varepsilon$ can be characterized by the effective electron
temperature,
\begin{equation}
f_{l}\left(  \varepsilon\right)  =N_{l}\frac{2\pi l_{B}^{2}}{AZ_{\parallel}%
}e^{-\varepsilon/T_{e}}, \label{e2}%
\end{equation}
where $l_{B}^{2}=\hbar c/eB$, and $A$ is the surface area. According to the
normalization condition $\int f_{l}\left(  \varepsilon\right)  D_{l}\left(
\varepsilon\right)  d\varepsilon=N_{l}$ [here $D_{l}\left(  \varepsilon
\right)  $ is the density-of-state function for the corresponding subband],
$Z_{\Vert}=\sum_{n}e^{-\varepsilon_{n}/T_{e}}$.

The approach reported here will be formulated in a quite general way to
be applicable for any weak quasi-elastic scattering. As important examples, we
shall consider interactions which are well established for SEs on liquid
helium. Vapor atoms are described by a free-particle energy spectrum
$\varepsilon_{\mathbf{K}}^{\left(  a\right)  }=\hbar^{2}K^{2}/2M$ with $M\gg
m_{e}$. For electron interaction with vapor atoms, it is conventional to adopt
the effective potential approximation
\begin{equation}
H_{int}^{\left(  a\right)  }=V^{\left(  a\right)  }\sum_{e}\sum_{a}%
\delta\left(  \mathbf{R}_{e}-\mathbf{R}_{a}\right)  , \label{e3}%
\end{equation}
where $V^{\left(  a\right)  }$ is proportional to the electron-atom scattering
length~\cite{SaiAok-78}. Ripplons represent a sort of 2D phonons, and the
electron-ripplon interaction Hamiltonian is usually written as
\begin{equation}
H_{int}^{(r)}=\frac{1}{\sqrt{A}}\sum_{e}\sum_{\mathbf{q}}U_{q}(z_{e}%
)Q_{q}\left(  b_{\mathbf{q}}+b_{-\mathbf{q}}^{\dag}\right)
e^{i\mathbf{q}\cdot\mathbf{r} _{e}}, \label{e4}%
\end{equation}
where $Q_{q}=\sqrt{\hbar q/2\rho\omega_{q}}$, $\omega_{q}\simeq\sqrt
{\alpha/\rho}q^{3/2}$ is the ripplon spectrum,
$\mathbf{R}_{e}=\left\{  z_{e},\mathbf{r}_{e}\right\}  $, $\hbar\mathbf{q}$ is
the ripplon momentum, $b_{-\mathbf{q}}^{\dag}$ and $b_{\mathbf{q}}$ are the
creation and destruction operators, and $U_{q}(z_{e})$ is the electron-ripplon
coupling~\cite{MonKon-04} which has a complicated dependence on $q$.

For both kinds of SE scattering, the energy exchange at
a collision is extremely small, which allows to consider scattering events as
quasi-elastic processes. In the case of vapor atoms, it is so because $M\gg m_{e}$.
One-ripplon scattering processes are quasi-elastic ($\hbar\omega_{q}\ll T$)
because the wave-vector of a ripplon involved is usually
restricted by the condition $ql_{B}\lesssim1$.

For quasi-elastic processes in a 2D electron system under magnetic field,
probabilities of electron scattering are usually found in the self-consistent
Born approximation (SCBA)~\cite{AndUem-74}. Following Ref.~\onlinecite{KubMiyHas-65},
we shall express the scattering probabilities in terms of the level densities at the
initial and the final states. Then, Landau level densities will be broadened
according to the SCBA~\cite{AndUem-74} or to the cumulant expansion
method~\cite{Ger-76},
\begin{equation}
D_{l}\left(  \varepsilon\right)  =-\frac{A}{2\pi^{2}l_{B}^{2}\hbar}\sum
_{n}\operatorname{Im}G_{l,n}\left(  \varepsilon\right)  , \label{e5}%
\end{equation}
where $G_{l,n}\left(  \varepsilon\right)  $ is the single-electron Green's
function. The later method is a bit more convenient for analytical evaluations
because it results in a Gaussian shape of level densities%
\begin{equation}
-\operatorname{Im}G_{l,n}\left(  \varepsilon\right)  =\frac{\sqrt{2\pi}\hbar
}{\Gamma_{l,n}}\exp\left[  -\frac{2\left(  \varepsilon-\varepsilon_{n}\right)
^{2}}{\Gamma_{l,n}^{2}}\right]  . \label{e6}%
\end{equation}
Here $\Gamma_{l,n}$ coincides with the broadening of Landau levels
given in the SCBA. For different scattering regimes of SEs, equations for $\Gamma_{l,n}$
are given in Ref.~\onlinecite{MonKon-04}. We shall also take into account an additional
increase in $\Gamma_{2,n}$ due to inter-subband scattering.

Effects considered in this work are important only under the condition
$\Gamma_{l,n}\ll T$ which is fulfilled for
SEs on liquid helium. Therefore, we shall disregard small corrections to
$Z_{\Vert}$ caused by collision broadening because they are proportional to
$\Gamma_{l,n}^{2}/8T_{e}^{2}$. In other equations, sometimes
we shall keep terms proportional to $\Gamma_{l,n}/T_{e}$, if they provide
important physical properties.

Average scattering probabilities of SEs on liquid helium and even the
effective collision frequency $\nu_{\mathrm{eff}}$ can be expressed in terms
of the dynamical structure factor (DSF) of the 2D electron liquid~\cite{MonKon-04} $S\left(
q,\omega\right)  $. This procedure somehow reminds the
theory of thermal neutron (or X-ray) scattering by solids, where the
scattering cross-section is expressed as an integral form of a DSF. Without MW
irradiation, most of unusual properties of the quantum magnetotransport of SEs
on liquid helium are well described by the equilibrium
DSF of the 2D electron liquid~\cite{MonTesWyd-02,MonKon-04}. A multisubband electron
system is actually a set of 2D electron systems. Therefore, the single factor
$S\left(  q,\omega\right)  $ is not appropriate for description of
inter-subband electron scattering. Luckily, for non-interacting electrons, we
can easily find an extension of $S\left(  q,\omega\right)  $ which could be
used in expressions for average scattering probabilities of a multisubband
system:%
\begin{eqnarray}
S_{l,l^{\prime}}\left(  q,\omega\right)  =\frac{2}{\pi\hbar Z_{\parallel}}%
\sum_{n,n^{\prime}}\int d\varepsilon e^{-\varepsilon/T_{e}}J_{n,n^{\prime}%
}^{2}(x_{q}) \times \nonumber \\
\times
\operatorname{Im}G_{l,n}\left(  \varepsilon\right)
\operatorname{Im}G_{l^{\prime},n^{\prime}}\left(  \varepsilon+\hbar
\omega\right)  , \label{e7}%
\end{eqnarray}
where%
\[
J_{n,n^{\prime}}^{2}(x)=\frac{[\min(n,n^{\prime})]!}{[\max(n,n^{\prime
})]!}x^{|n-n^{\prime}|}e^{-x}\left[  L_{\min(n,n^{\prime})}^{|n-n^{\prime}%
|}(x)\right]  ^{2},
\]
$x_{q}=q^{2}l_{B}^{2}/2$, and $L_{n}^{m}(x)$ are the associated Laguerre
polynomials. The factor $S_{l,l^{\prime}}\left(  q,\omega\right)  $ contains
the level densities at the initial and the final states, and it includes
averaging over initial in-plane states. At $l=l^{\prime}$,
this factor coincides with the DSF of a
nondegenerate 2D system of non-interacting electrons. Generally, $S_{l,l^{\prime}%
}\left(  q,\omega\right)  $ is not the dynamical structure factor of the whole
system, nevertheless this function is very useful for description
of dissipative processes in presence of MW irradiation.

As a useful example, consider the average inter-subband scattering rate
$\bar{\nu}_{l\rightarrow l^{\prime}}$ caused by quasi-elastic scattering, which
is important for obtaining subband occupancies $\bar{n}_{l}=N_{l}/N_{e}$
under the MW resonance~\cite{MonSokStu-10}. Using the damping theoretical
formulation~\cite{KubMiyHas-65} and the SCBA~\cite{AndUem-74}, $\bar{\nu}_{l\rightarrow
l^{\prime}}$ can be represented in the following form%
\begin{equation}
\ \bar{\nu}_{l\rightarrow l^{\prime}}=\frac{\hbar}{m_{e}A}\sum_{\mathbf{q}%
}\chi_{l,l^{\prime}}\left(  q\right)  S_{l,l^{\prime}}\left(  q,\omega
_{l,l^{\prime}}\right)  , \label{e8}%
\end{equation}
where $\chi_{l,l^{\prime}}$($=\chi_{l^{\prime},l}$) describes electron
coupling with scatterers. For SEs on liquid helium, we have two kinds of
scatterers: ripplons and vapor atoms. Therefore, $\chi_{l,l^{\prime}}%
=\chi_{l,l^{\prime}}^{\left(  r\right)  }+\chi_{l,l^{\prime}}^{\left(
a\right)  }$. Electron-ripplon scattering gives
\begin{equation}
\chi_{l,l^{\prime}}^{\left(  r\right)  }\left(  q\right)  =\frac{m_{e}}%
{\hbar^{3}}Q_{q}^{2}2N_{q}\left\vert \left(  U_{q}\right)  _{l,l^{\prime}%
}\right\vert ^{2}\simeq\frac{m_{e}T}{\alpha\hbar^{3}q^{2}}\left\vert \left(
U_{q}\right)  _{l,l^{\prime}}\right\vert ^{2},\text{ \ } \label{e9}%
\end{equation}
where $N_{q}=\left(  e^{\hbar\omega_{q}/T}-1\right)  ^{-1}\gg1$, and $\left(
U_{q}\right)  _{l,l^{\prime}}\equiv\left\langle l\right\vert U_{q}\left(
z_{e}\right)  \left\vert l^{\prime}\right\rangle $. For electron scattering at
vapor atoms,%
\begin{equation}
\chi_{l,l^{\prime}}^{\left(  a\right)  }\left(  q\right)  =\nu_{0}%
^{(a)}p_{l,l^{\prime}},\text{ \ \ \ }\nu_{0}^{(a)}=\frac{m_{e}n_{a}^{\left(
3D\right)  }\left( V^{(a)}\right) ^{2}}{\hbar^{3}B_{1,1}} \label{e10}%
\end{equation}
where%
\[
\text{\ }p_{l,l^{\prime}}=\frac{B_{1,1}}{B_{l,l^{\prime}}},\text{
\ \ \ }B_{l,l^{\prime}}^{-1}=L_{z}^{-1}\sum_{K_{z}}\left\vert \left(
e^{iK_{z}z_{e}}\right)  _{l^{\prime},l}\right\vert ^{2},
\]
$L_{z}$ is the height above the liquid surface, $n_{a}^{\left(  3D\right)  }$
is the density of vapor atoms, and $K_{z}$ is the projection of the vapor atom
wave-vector. The $\nu_{0}^{(a)}$ represents the SE collision frequency at
vapor atoms for $B=0$.

The generalized factor of a multi-subband 2D electron system $S_{l,l^{\prime}%
}\left(  q,\omega \right)  $ will be used throughout this work
because its basic property
\begin{equation}
S_{l,l^{\prime}}\left(  q,-\omega\right)  =e^{-\hbar\omega/T_{e}}S_{l^{\prime
},l}\left(  q,\omega\right)  \label{e11}%
\end{equation}
allows us straightforwardly to obtain terms responsible for negative
dissipation. This property follows from the detailed balancing for
quasi-elastic processes, $\bar{\nu}_{l^{\prime}\rightarrow l}=e^{-\Delta
_{l,l^{\prime}}/T_{e}}\bar{\nu}_{l\rightarrow l^{\prime}}$, and also directly
from the definition of Eq.~(\ref{e7}). Using Gaussian level shapes of
Eq.~(\ref{e6}), one can find%
\begin{equation}
S_{l,l^{\prime}}\left(  q,\omega\right)  =\frac{2\pi^{1/2}\hbar}{Z_{\parallel
}}\sum_{n,n^{\prime}}\frac{J^{2}_{n,n^{\prime}}(x_{q}) %
}{\Gamma_{l,n;l^{\prime},n^{\prime}}}e^{-\varepsilon_{n}/T_{e}}%
I_{l,n;l^{\prime},n^{\prime}}\left(  \omega\right)  , \label{e12}%
\end{equation}
where $2\Gamma_{l,n;l^{\prime},n^{\prime}}^{2}=\Gamma_{l,n}^{2}+\Gamma
_{l^{\prime},n^{\prime}}^{2}$, and
\begin{eqnarray}
I_{l,n;l^{\prime},n+m}\left(  \omega\right)  =
\nonumber \,\,\,\,\,\,\,\,\,\,\,\,\,\,\,\,\,\,
\,\,\,\,\,\,\,\,\,\,\,\,\,\,\,\,\,\,\\\
=\exp\left[
-\left( \frac{\hbar \omega -m\hbar \omega _{c}%
-\Gamma_{l,n}^{2}/4T_{e}}{\Gamma_{l,n;l^{\prime},n+m}} \right) ^{2}+\frac{\Gamma_{l,n}^{2}}{8T_{e}^{2}}\right]. \label{e13}%
\end{eqnarray}
The $S_{l,l^{\prime}}\left(  q,\omega\right)  $, as a
function of frequency, has sharp maxima when $\omega$ \ equals the in-plane
excitation energy $\left(  n^{\prime}-n\right)  \hbar\omega_{c}$. The
parameter $\Gamma_{l,n;l^{\prime},n^{\prime}}^{2}$ describes broadening of
these maxima. Eqs.~(\ref{e12}) and (\ref{e13}) satisfy the condition of
Eq.~(\ref{e11}). Terms of the order of $\left(  \Gamma_{l,n}/T_{e}\right)
^{2}$ entering the argument of Eq.~(\ref{e13}) could be omitted, as it was done
for $Z_{\parallel}$, because even the linear in $\Gamma_{l,n}/T_{e}$ term
provides us the necessary condition of Eq.~(\ref{e11}). Anyway, our final
results will be represented in forms
which allow to disregard even the linear in $\Gamma_{l,n}/T_{e}$
term entering $I_{l,n;l^{\prime},n^{\prime}}\left(  \omega\right)  $. \

Consider the decay rate of the first excited subband $\bar{\nu}_{2\rightarrow
1}$. Under typical experimental condition, $\Gamma_{l,n;l^{\prime}%
,n^{\prime}}$ is much smaller than $\hbar\omega_{c}$. Therefore, most of terms
entering $S_{l,l^{\prime}}\left(  q,\omega_{2,1}\right)  $ are exponentially
small and can be disregarded. The exceptional terms satisfy the condition
$n^{\prime}-n=m^{\ast}\left(  B\right)  $,
where $m^{\ast}\left(  B\right)  \equiv\mathrm{round}\left(  \omega
_{2,1}/\omega_{c}\right)  $ is an integer nearest to $\omega_{2,1}/\omega_{c}%
$. In this notation,
\begin{eqnarray}
\text{\ }\bar{\nu}_{2\rightarrow1}=\sum_{n=0}^{\infty}\frac{e^{-\varepsilon
_{n}/T_{e}}}{Z_{\parallel}}\frac{\hbar\omega_{c}\beta_{2,n;1,n+m^{\ast}}}%
{\pi^{1/2}\Gamma_{2,n;1,n+m^{\ast}}}\times \nonumber \\
\times \exp\left\{  -\frac{\hbar^{2}\left(
\omega_{2,1}-m^{\ast}\omega_{c}\right)  ^{2}}{\Gamma_{2,n;1,n+m^{\ast}}^{2}%
}\right\}  , \label{e14}%
\end{eqnarray}
where%
\[
\beta_{l,n;l^{\prime},n+m}=\int_{0}^{\infty}\chi_{l,l^{\prime}}J_{n,n+m}%
^{2}(x_{q})dx_{q}.
\]
For electron scattering at vapor atoms, $\beta_{l,n;l^{\prime},n+m}\equiv
\beta_{l,l^{\prime}}^{\left(  a\right)  }=\nu_{0}^{(a)}p_{l,l^{\prime}}$ which
coincides with$\ \chi_{l,l^{\prime}}^{\left(  a\right)  }$. In the case of
electron-ripplon scattering, $\beta_{2,n;1,n+m}$ has a more complicated
expression due to a particular form of $U_{q}(z_{e})$ entering the definition
of $\chi_{l,l^{\prime}}^{\left(  r\right)  }$. The $\bar{\nu}_{2\rightarrow
1}(B)$ is a 1/B-periodic function. It has sharp maxima when $\omega
_{2,1}/\omega_{c}$ equals an integer. In the argument of the exponential
function of Eq.~(\ref{e14}), we have disregarded terms which are small for
$\Gamma_{l,n}/T_{e}\ll1$.

Transition rates $\bar{\nu}_{l\rightarrow l^{\prime}}$ determine subband
occupancies $\bar{n}_{l}$ under the MW resonance. At low electron
temperatures, the two-subband model is applicable, and the rate equation gives%
\begin{equation}
\frac{\bar{n}_{2}}{\bar{n}_{1}}=\frac{r+e^{-\Delta_{2,1}/T_{e}}\bar{\nu
}_{2\rightarrow1}}{r+\bar{\nu}_{2\rightarrow1}},\text{ \ } %
\label{e15}%
\end{equation}
where $r$ is the stimulated absorption (emission) rate due to the MW field,
and $\bar{n}_{1}+\bar{n}_{2}=1$. Thus, under the MW resonance,
magnetooscillations of $\bar{\nu}_{2\rightarrow1}$ lead to magnetooscillations
of subband occupancies $\bar{n}_{1}$ and $\bar{n}_{2}$. For further analysis,
it is important that MW excitation provides the condition $\bar{n}_{2}>\bar
{n}_{1}e^{-\Delta_{2,1}/T_{e}}$, which is the main cause of negative
momentum dissipation.

\section{Magnetoconductivity under resonance MW irradiation}

Consider now an infinite isotropic multisubband 2D electron system under an in-plane
dc-electric field, assuming arbitrary occupancies of surface subbands $\bar{n}_{l}$
induced by the MW resonance. In the linear transport regime,
the average friction force acting on electrons
due to interaction with scatterers $\mathbf{F}_{\mathrm{scat}}$ is
proportional to the average electron velocity $\mathbf{V}_{\mathrm{av}%
}=\left\langle \mathbf{v}_{e}\right\rangle $. This relationship can be
conveniently written as $\mathbf{F}_{\mathrm{scat}}=-N_{e}m_{e}\nu
_{\mathrm{eff}}\mathbf{V}_{\mathrm{av}}$, where the proportionality factor
$\nu_{\mathrm{eff}}$ represents an effective collision frequency which depends
on $B$ and, generally, on electron density. The $\mathbf{F}_{\mathrm{scat}}$
is balanced by the average Lorentz force $\left\langle \mathbf{F}%
_{\mathrm{field}}\right\rangle $, which yields the usual Drude form for the
electron conductivity tensor $\sigma_{i,k}$, where the quasi-classical
collision frequency $\nu_{0}$ is substituted for $\nu_{\mathrm{eff}}%
$~\cite{MonTesWyd-02,MonKon-04}.

The effective collision frequency $\nu_{\mathrm{eff}}$ can be obtained
directly from the expression for the average momentum gained by scatterers per
unite time. Usually, to describe momentum relaxation, one have to obtain
deviations of the in-plane electron distribution function from the simple form
of Eq.~(\ref{e2}) induced by the dc-electric field. For the highly
correlated 2D system of SEs on liquid helium under magnetic field, this
problem was solved in a general way, assuming that in the
center-of-mass reference frame the electron DSF has its
equilibrium form $S^{\left(  0\right)  }\left(  q,\omega\right)  $.
In the laboratory frame, its frequency
argument acquires the Doppler shift $S\left(  \mathbf{q},\omega\right)
=S^{\left(  0\right)  }\left(  q,\omega-\mathbf{q\cdot V}_{\mathrm{av}%
}\right)  $ due to Galilean invariance~\cite{MonTesWyd-02,MonKon-04}. This
approach is similar to the description of electron transport by a velocity
shifted Fermi-function of the kinetic equation method, where $\mathbf{V}%
_{\mathrm{av}}$ is found from the momentum balance equation. The same
properties can be ascribed to the generalized factor~\cite{Mon-11} $S_{l,l^{\prime}}\left(
\mathbf{q},\omega\right)  $.

Here, we consider a different way,
taking into account that $\mathbf{F}_{\mathrm{scat}}$, as well as the momentum
gained by scatterers, can be evaluated in any inertial reference frame. We
choose the reference frame fixed to the electron liquid center-of-mass,
because in it the in-plane distribution function of highly correlated
electrons ($\nu_{ee}\gg\nu_{\mathrm{eff}}$) has its simplest form of
Eq.~(\ref{e2}), and the generalized factor $S_{l,l^{\prime}}\left(
\mathbf{q},\omega\right)  $ has it equilibrium form of Eq.~(\ref{e7}). Then,
$\mathbf{F}_{\mathrm{scat}}$ can be considered as the drag due to moving
scatterers. At the same time, distribution functions of scatterers which are
not affected by external fields can be easily found according to well-known rules.

In the electron liquid center-of-mass reference frame, the in-plane spectrum
of electrons is close to the Landau spectrum, because the driving electric
field $\mathbf{E}^{\prime}=\mathbf{E}-(1/c)\mathbf{B\times V}_{\mathrm{av}}$
is nearly zero, at least for $\omega_{c}\gg\nu_{\mathrm{eff}}$. At the same
time, in this frame the ripplon excitation energy is changed to $E_{\mathbf{q}%
}^{\left(  r\right)  }=\hbar\omega_{\mathbf{q}}-\hbar\mathbf{qV}_{\mathrm{av}%
}$, because the gas of ripplons moves as a whole with the drift velocity equal
to $-\mathbf{V}_{\mathrm{av}}$. The same Doppler shift correction
$-\hbar\mathbf{qV}_{\mathrm{av}}$ appears for the energy exchange in the case
of electron scattering at vapor atoms, even for the limiting case
$M\rightarrow\infty$ (impurities which are motionless in the laboratory reference frame).
In the electron center-of-mass reference frame, vapor atoms move with the velocity
$-\mathbf{V}_{\mathrm{av} }$ and hit electrons which results in the energy exchange
$-\hbar \mathbf{qV}_{\mathrm{av}}$.

Describing electron-ripplon scattering probabilities in terms of the
equilibrium factor $S_{l,l^{\prime}}\left(  q,\omega\right)  $, as discussed
above, contributions to the frictional force from creation and destruction
processes can be found as
\begin{widetext}
\begin{equation}
\mathbf{F}_{\mathrm{scat}}=-\frac{N_{e}}{\hbar^{2}A}\sum_{\mathbf{q}}%
\hbar\mathbf{q}Q_{q}^{2}\sum_{l,l^{\prime}}\bar{n}_{l}\left\vert \left(
U_{q}\right)  _{l,l^{\prime}}\right\vert ^{2}
\left[  \left(  N_{\mathbf{q}}+1\right)  S_{l,l^{\prime}}\left(
q,\omega_{l,l^{\prime}}-E_{\mathbf{q}}^{\left(  r\right)  }/\hbar\right)
-N_{\mathbf{q}}S_{l,l^{\prime}}\left(  q,\omega_{l,l^{\prime}}+E_{\mathbf{q}%
}^{\left(  r\right)  }/\hbar\right)  \right]  . \label{e16}%
\end{equation}
\end{widetext}
It is clear that disregarding the Doppler-shift correction $-\hbar
\mathbf{qV}_{\mathrm{av}}$ in this equation yields zero result. This
correction enters the ripplon distribution function $N_{\mathbf{q}}$ and the
frequency argument of the factor $S_{l,l^{\prime}}$. In the linear transport
regime, the Doppler-shift correction entering the ripplon distribution
function is unimportant. This can be seen directly from Eq.~(\ref{e16}):
setting $E_{\mathbf{q}}^{\left(  r\right)  }\rightarrow0$ in the frequency
argument of $S_{l,l^{\prime}}$ gives zero result for $\mathbf{F}_{\mathrm{scat}}$.
Therefore, in this equation one can substitute
$N_{\mathbf{q}}$ for $N_{q}$ defined in Eq.~(\ref{e9}). We can also disregard
$\hbar\omega_{q}$ in the frequency argument of $S_{l,l^{\prime}}$. Then,
interchanging the running indices of the second term in the square brackets,
and using the basic property of $S_{l,l^{\prime}}\left(  q,\omega\right)  $
given in Eq.~(\ref{e11}), Eq.~(\ref{e16}) can be represented as%
\begin{eqnarray}
\mathbf{F}_{\mathrm{scat}}=-\frac{N_{e}\hbar}{2m_{e}A}\sum_{l,l^{\prime}%
}\sum_{\mathbf{q}}\hbar\mathbf{q}\chi_{l,l^{\prime}}\left(  q\right)
S_{l,l^{\prime}}\left(  q,\omega_{l,l^{\prime}}+\mathbf{q\cdot
V}_{\mathrm{av}}\right)
\nonumber \\
\times   \left(  \bar{n}_{l}-\bar{n}_{l^{\prime}}%
e^{-\Delta_{l,l^{\prime}}/T_{e}} e^{-\hbar\mathbf{q\cdot V}_{\mathrm{av}}/T_{e}}\right), \,\,\,\,\,\,\,\,\,\,\,
\,\,\,\,\,\,\,\,\,\,  \label{e17}%
\end{eqnarray}
where $\chi_{l,l^{\prime}}\left(  q\right)  =\chi_{l,l^{\prime}}^{\left(
r\right)  }\left(  q\right)  $ was defined in Eq.~(\ref{e9}). This equation has
the most convenient form for expansion in $\hbar\mathbf{q\cdot V}%
_{\mathrm{av}}$.

A similar equation for $\mathbf{F}_{\mathrm{scat}}$ can be found considering
electron scattering at vapor atoms. Evaluating momentum relaxation rate, one
can disregard $\hbar\varkappa_{\mathbf{K,K}^{\prime}}=\varepsilon
_{\mathbf{K}^{\prime}-\mathbf{K}}^{\left(  a\right)  }-\varepsilon
_{\mathbf{K}^{\prime}}^{\left(  a\right)  }$ which represents the energy
exchange at a collision in the laboratory reference frame. In the
center-of-mass reference frame, Doppler-shift corrections enter the vapor atom
distribution function $N_{\mathbf{K}^{\prime}}^{\left(  a\right)  }$ and the
frequency argument of the factor $S_{l,l^{\prime}}$ due to the new energy
exchange at a collision $-\hbar\mathbf{qV}_{\mathrm{av}}$. The correction
entering\ $N_{\mathbf{K}^{\prime}}^{\left(  a\right)  }$ is unimportant
because of the normalization condition: $\sum_{\mathbf{K}^{\prime}%
}N_{\mathbf{K}^{\prime}}^{\left(  a\right)  }=$ $n_{a}^{\left(  3D\right)
}L_{z}A$. Therefore, we have
\begin{eqnarray}
\mathbf{F}_{\mathrm{scat}}=-\frac{N_{e}\hbar\nu_{0}^{(a)}}{m_{e}A}%
\sum_{l,l^{\prime}}\bar{n}_{l}p_{l,l^{\prime}}\times \nonumber \\
\times \sum_{\mathbf{q}}\hbar
\mathbf{q}S_{l,l^{\prime}}\left(  q,\omega_{l,l^{\prime}}+\mathbf{q\cdot
V}_{\mathrm{av}}\right)  . \label{e18}%
\end{eqnarray}
In order to obtain the form of Eq.~(\ref{e17}), we represent the right side of
Eq.~(\ref{e18}) as a sum of two identical halves and change the running indices
in the second half: $\mathbf{q\rightarrow-q}$ and $l\rightleftarrows
l^{\prime}$. Then, the basic property of $S_{l,l^{\prime}}\left(
q,\omega\right)  $ yields Eq.~(\ref{e17}) \ with $\chi_{l,l^{\prime}}%
=\chi_{l,l^{\prime}}^{\left(  a\right)  }$, where $\chi_{l,l^{\prime}%
}^{\left(  a\right)  }$ is from Eq.~(\ref{e10}).

Thus, Eq.~(\ref{e17}) is applicable for both scattering mechanisms. In the
general case, $\chi_{l,l^{\prime}}=\chi_{l,l^{\prime}}^{\left(  r\right)
}+\chi_{l,l^{\prime}}^{\left(  a\right)  }$. The effective collision frequency
under magnetic field $\nu_{\mathrm{eff}}$ can be found expanding
Eq.~(\ref{e17}) in $\mathbf{q\cdot V}_{\mathrm{av}}$ up to linear terms. We
shall represent $\nu_{\mathrm{eff}}$ as a sum of two different contributions:
$\nu_{\mathrm{eff}}=\nu_{\mathrm{N}}+\nu_{\mathrm{A}}$. The normal
contribution $\nu_{\mathrm{N}}$ originates from the expansion of the
exponential factor $\exp\left(  -\hbar\mathbf{q\cdot V}_{\mathrm{av}}%
/T_{e}\right)  $. In turn, $\nu_{\mathrm{N}}$\ can be represented as a sum of
contributions from intra-subband and inter-subband scattering $\nu
_{\mathrm{N}}=\nu_{\mathrm{N},intra}+\nu_{\mathrm{N},inter}$. The sums of
$\nu_{\mathrm{N},inter}$ take account of all $l,l^{\prime}$. It is useful to
rearrange terms with $l<l^{\prime}$ ($\Delta_{l,l^{\prime}}<0$) by
interchanging the running indices $l\rightleftarrows l^{\prime}$, and using
the basic property of $S_{l,l^{\prime}}\left(  q,\omega \right)
$. Then, we have
\begin{equation}
\nu_{\mathrm{N},intra}=\frac{\hbar\omega_{c}^{2}}{4\pi T_{e}}\sum_{l}\bar
{n}_{l}\int_{0}^{\infty}x_{q}\chi_{l,l}\left(  q\right)  S_{l,l}\left(
q,0\right)  dx_{q}, \label{e19}%
\end{equation}%
\begin{eqnarray}
\nu_{\mathrm{N},inter}=\frac{\hbar\omega_{c}^{2}}{4\pi T_{e}}\sum
_{l>l^{\prime}}\left(  \bar{n}_{l}+\bar{n}_{l^{\prime}}e^{-\Delta
_{l,l^{\prime}}/T_{e}}\right)  \times \,\,\,\,\,\,\,\,\,\,\, \,\,\,\,\,
\,\,\,\,\,\,\,\,\,\,\, \nonumber \\
\times \int_{0}^{\infty}x_{q}\chi_{l,l^{\prime}%
}\left(  q\right)  S_{l,l^{\prime}}\left(  q,\omega_{l,l^{\prime}}\right)
dx_{q}. \,\,\,\,\,\,\,\,\,\,\,\,\, \label{e20}%
\end{eqnarray}
The $\nu_{\mathrm{N}}\left(  B\right)  $ is always positive. In the limiting
case of a one-subband 2D electron system ($\bar{n}_{l}=\delta_{l,1}$),
Eq.~(\ref{e19}) reproduces the known relationship between the effective
collision frequency and the electron DSF~\cite{MonKon-04}. In the parentheses
of Eq.~(\ref{e20}), the first term is due to scattering from $l$ to $l^{\prime
}$, while the second term describes the contribution of scattering back from
$l^{\prime}$ to $l$. It should be noted that the forms of Eqs.~(\ref{e19}) and
(\ref{e20}) allow to simplify $S_{l,l^{\prime}}\left(  q,\omega_{l,l^{\prime}%
}\right)  $ of Eq.~(\ref{e12}) by disregarding small corrections proportional
to $\Gamma_{l,n}/T_{e}$ and $\left(  \Gamma_{l,n}/T_{e}\right)  ^{2}$ entering
$I_{l,n;l^{\prime},n^{\prime}}\left(  \omega\right)  $ defined by
Eq.~(\ref{e13}).

The anomalous contribution to the effective collision frequency $\nu
_{\mathrm{A}}\left(  B\right)  $ can be found from Eq.~(\ref{e17}) expanding
$S_{l,l^{\prime}}\left(  q,\omega_{l,l^{\prime}}+\mathbf{q\cdot V}%
_{\mathrm{av}}\right)  $ in $\mathbf{q\cdot V}_{\mathrm{av}}$, and setting
$\exp\left(  -\hbar\mathbf{q\cdot V}_{\mathrm{av}}/T_{e}\right)  \rightarrow1$
in the parentheses. In this case, to rearrange terms with $l<l^{\prime}$
($\Delta_{l,l^{\prime}}<0$), we shall use the property
\[
S_{l^{\prime},l}^{\prime}\left(  q,-\omega\right)  =-e^{-\frac{\hbar\omega
}{T_{e}}}S_{l,l^{\prime}}^{\prime}\left(  q,\omega\right)  +\frac{\hbar}%
{T_{e}}e^{-\frac{\hbar\omega}{T_{e}}}S_{l,l^{\prime}}\left(  q,\omega\right)
\simeq
\]%
\begin{equation}
\simeq-e^{-\frac{\hbar\omega}{T_{e}}}S_{l,l^{\prime}}^{\prime}\left(
q,\omega\right)  . \label{e21}%
\end{equation}
Here $S_{l,l^{\prime}}^{\prime}\left(  q,\omega\right)  \equiv\partial
S_{l,l^{\prime}}\left(  q,\omega\right)  /\partial\omega$, and the last
transformation assumes that $\Gamma_{l,n}\ll T_{e}$. Interchanging the running
indices $l\rightleftarrows l^{\prime}$ of terms with $\Delta_{l,l^{\prime}}<0$
and using Eq.~(\ref{e21}), $\nu_{\mathrm{A}}\left(  B\right)  $ can be found
as
\begin{eqnarray}
\text{\ \ \ }\nu_{\mathrm{A}}=\frac{\omega_{c}^{2}}{2\pi}\sum_{l>l^{\prime}%
}\left(  \bar{n}_{l}-\bar{n}_{l^{\prime}}e^{-\Delta_{l,l^{\prime}}/T_{e}%
}\right)  \times \nonumber \\
\times \int_{0}^{\infty}x_{q}\chi_{l,l^{\prime}}\left(  q\right)
S_{l,l^{\prime}}^{\prime}\left(  q,\omega_{l,l^{\prime}}\right)  dx_{q}.
\label{e22}%
\end{eqnarray}
As compared to $\nu_{\mathrm{N},inter}$ of the normal contribution, here the
second term in parentheses has the opposite sign. Therefore, for usual
Boltzmann distribution of subband occupancies, $\nu_{\mathrm{A}}\left(
B\right)  =0$. The anomalous contribution appears only when $\bar{n}_{l}%
\neq\bar{n}_{l^{\prime}}e^{-\Delta_{l,l^{\prime}}/T_{e}}$, which occurs under
the MW resonance condition $\omega=\omega_{l,l^{\prime}}$.

In the form of Eq.~(\ref{e22}), it is possible to use a simplified expression%
\begin{widetext}
\begin{equation}
S_{l,l^{\prime}}^{\prime}\left(  q,\omega_{l,l^{\prime}}\right)  \simeq
-\frac{2\pi^{1/2}\hbar}{Z_{\parallel}}\sum_{n,n^{\prime}}\frac{
J^{2}_{n,n^{\prime}}(x_{q})}{\Gamma_{l,n;l^{\prime},n^{\prime}}%
}e^{-\varepsilon_{n}/T_{e}}
\exp\left\{  -\frac{\left[  \Delta_{l,l^{\prime}}-\left(  n^{\prime
}-n\right)  \hbar\omega_{c}\right]  ^{2}}{\Gamma_{l,n;l^{\prime},n^{\prime}%
}^{2}}\right\}  \frac{2\hbar^{2}\left[  \omega_{l,l^{\prime}}-\left(
n^{\prime}-n\right)  \omega_{c}\right]  }{\Gamma_{l,n;l^{\prime},n^{\prime}%
}^{2}}, \label{e23}%
\end{equation}
\end{widetext}
which disregards terms proportional to $\Gamma_{l,n}/T_{e}$ and $\left(
\Gamma_{l,n}/T_{e}\right)  ^{2}$. From Eqs.~(\ref{e22}) and (\ref{e23}) one can
see that at $\bar{n}_{l}>\bar{n}_{l^{\prime}}e^{-\Delta_{l,l^{\prime}}/T_{e}}%
$, the sign of $\nu_{\mathrm{A}}\left(  B\right)  $ is opposite to the sign of
$\omega_{l,l^{\prime}}-\left(  n^{\prime}-n\right)  \omega_{c}$. Therefore,
$\nu_{\mathrm{A}}\left(  B\right)  <0$ when the magnetic field $B$ is slightly
lower the commensurability condition $\Delta_{2,1}/\hbar\omega_{c}=m$ (here $m$
is an integer), which agrees with the experimental observation for minima of
$\sigma_{xx}$.

For further analysis, it is convenient to introduce%
\begin{equation}
\text{\ \ }\lambda_{l,n;l^{\prime},n^{\prime}}=\int_{0}^{\infty}x_{q}%
\chi_{l,l^{\prime}}\left(  q\right)  J_{n,n^{\prime}}^{2}\left(  x_{q}\right)
dx_{q}\text{ .\ \ } \label{e24}%
\end{equation}
When referring to a particular scattering mechanism, we shall use a superscript, $\lambda_{l,n;l^{\prime},n^{\prime}}=\lambda_{l,n;l^{\prime},n^{\prime
}}^{\left(  r\right)  }+\lambda_{l,n;l^{\prime},n^{\prime}}^{\left(  a\right)
}$. Consider a two-subband model which is valid at low enough electron
temperatures. Using the new definitions given above, the normal contribution
to the effective collision frequency can be represented as
\begin{equation}
\nu_{\mathrm{N},intra}=\sum_{n=0}^{\infty}\frac{e^{-\varepsilon_{n}/T_{e}%
}\left(  \hbar\omega_{c}\right)  ^{2}}{2\sqrt{\pi}T_{e}Z_{\parallel}}\left[
\bar{n}_{1}\frac{\lambda_{1,n;1,n}}{\Gamma_{1,n}}+\bar{n}_{2}\frac
{\lambda_{2,n;2,n}}{\Gamma_{2,n}}\right]  , \label{e25}%
\end{equation}%
\begin{eqnarray}
\nu_{\mathrm{N},inter}=\left(  \bar{n}_{2}+\bar{n}_{1}e^{-\Delta_{2,1}/T_{e}%
}\right) \frac{\left(  \hbar\omega_{c}\right)  ^{2}}{2\sqrt{\pi}T_{e}} \sum_{n=0}^{\infty}\frac{e^{-\varepsilon_{n}/T_{e}}}{Z_{\parallel
}}\times \nonumber \\
\times
\frac{\lambda_{2,n;1,n+m^{\ast}}%
}{\Gamma_{2,n;1,n+m^{\ast}}}
\exp\left[  -\frac{\hbar^{2}\left(  \omega_{2,1}-m^{\ast}\omega
_{c}\right)  ^{2}}{\Gamma_{2,n;1,n+m^{\ast}}^{2}}\right]  ,
\,\,\,\,\,\,\,%
\label{e26}%
\end{eqnarray}
where $m^{\ast}\equiv\mathrm{round}\left(  \omega_{2,1}/\omega_{c}\right)  $
is the function of $B$ defined in the previous Section. The $\nu_{\mathrm{N},intra}$ and
$\nu_{\mathrm{N},inter}$ have magnetooscillations of two kinds. Oscillations
of $\nu_{\mathrm{N},inter}$ are quite obvious, because quasi-elastic
inter-subband scattering increases sharply at the commensurability condition:
$\omega_{2,1}=m\omega_{c}$. The shape of these peaks is symmetrical with
respect to the point $\Delta_{2,1}/\hbar\omega_{c}=m$. It is formed by the
interplay of the exponential factor, having $\Gamma_{2,n;1,n+m^{\ast}}$ for
the broadening parameter, and the line-shapes of the subband occupancies. It
should be noted that at low electron temperatures, $\nu_{\mathrm{N},inter}$ is
exponentially small. The intra-subband scattering contribution $\nu
_{\mathrm{N},intra}$ oscillates with $1/B$ in an indirect way because of
oscillations in level occupancies $\bar{n}_{2}$ and $\bar{n}_{1}$ induced by
oscillations in the decay rate $\bar{\nu}_{2\rightarrow1}$, according to
Eqs.~(\ref{e14}) and (\ref{e15}). These oscillations have also a symmetrical
shape whose broadening is affected by the relation between $r$ and $\bar{\nu
}_{2\rightarrow1}$.

Magnetooscillations of $\nu_{\mathrm{A}}\left(  B\right)  $ have a completely
different shape:
\begin{eqnarray}
\text{\ }\nu_{\mathrm{A}}=-\left(  \bar{n}_{2}-\bar{n}_{1}e^{-\Delta
_{2,1}/T_{e}}\right) \frac{\left(  \hbar\omega_{c}\right)  ^{2}}{\sqrt{\pi}} \times \nonumber \,\,\,\,\,\,\,\,\,\,\,\,\,\, \\
\times \sum_{n=0}^{\infty}\frac{e^{-\varepsilon_{n}/T_{e}}%
}{Z_{\parallel}}\frac{\lambda
_{2,n;1,n+m^{\ast}}}{\Gamma_{2,n;1,n+m^{\ast}}^{2}}
\times \nonumber \,\,\,\,\,\,\,\,\,\,\,\,\,\, \\
\times \exp\left[  -\frac{\hbar^{2}\left(  \omega_{2,1}-m^{\ast}\omega
_{c}\right)  ^{2}}{\Gamma_{2,n;1,n+m^{\ast}}^{2}}\right]  \frac{2\hbar\left(
\omega_{2,1}-m^{\ast}\omega_{c}\right)  }{\Gamma_{2,n;1,n+m^{\ast}}},
\label{e27}%
\end{eqnarray}
In the ultra-quantum limit $\hbar\omega_{c}\gg T_{e}$, terms with $n>0$
entering Eq.~(\ref{e27}) can be omitted, which allows to describe
magneto-oscillations of $\nu_{\mathrm{A}}\left(  B\right)  $ in an analytical
form. In contrast with oscillations of the normal contribution $\nu
_{\mathrm{N}}$, in the vicinity of the commensurability condition,
$\nu_{\mathrm{A}}$ is an odd function of $\omega_{2,1}/\omega_{c}-m^{\ast}$.

Thus, the effective collision frequency $\nu_{\mathrm{eff}}=\nu_{\mathrm{N}}%
+\nu_{\mathrm{A}}$ and magnetoconductivity $\sigma_{xx}$ of SEs are found for
any given electron temperature. In order to obtain $T_{e}$ as a
function of the magnetic field, it is necessary to describe energy relaxation of SEs for
arbitrary subband occupancies.

\section{Energy dissipation}

It is instructive to analyze another important example of negative dissipation
which can be induced by the MW resonance. Consider the energy loss rate of a
multisubband 2D electron system due to interaction with scatterers. In this
case, there are no complications with the dc-driving electric field or with the Doppler
shifts which can be set to zero. This analysis will be important also for
description of electron heating due to decay of the SE state excited by the MW.

The energy loss rate per an electron due to one-ripplon creation and
destruction processes can be represented in terms of
$S_{l,l^{\prime}}\left(  q,\omega\right)  $ quite straightforwardly:%
\[
\dot{W}=-\frac{1}{\hbar^{2}A}\sum_{\mathbf{q}}\hbar\omega_{\mathbf{q}}%
Q_{q}^{2}\left(  N_{q}+1\right)  \sum_{l,l^{\prime}}\left\vert \left(
U_{q}\right)  _{l,l^{\prime}}\right\vert ^{2}\times
\]
\begin{equation}
\times\left[  \bar{n}_{l}S_{l,l^{\prime}}\left(  q,\omega_{l,l^{\prime}%
}-\omega_{q}\right)  -\bar{n}_{l}e^{-\hbar\omega_{q}/T}S_{l,l^{\prime}}\left(
q,\omega_{l,l^{\prime}}+\omega_{q}\right)  \right] .  \label{e28}
\end{equation}
Interchanging the running indices ($l,l^{\prime}$) in the second term, and
using the basic property of $S_{l,l^{\prime}}\left(  q,\omega\right)  $ given
in Eq.~(\ref{e11}), the terms entering the square brackets can be rearranged
as
\begin{equation}
S_{l,l^{\prime}}\left(  q,\omega_{l,l^{\prime}}-\omega_{q}\right)  \left[
\bar{n}_{l}-\bar{n}_{l^{\prime}}e^{-\Delta_{l,l^{\prime}}/T_{e}}%
e^{-\hbar\omega_{q}\left(  1/T-1/T_{e}\right)  }\right]  . %
\label{e29}%
\end{equation}
Since the processes considered here are quasi-elastic, we can expand this
equation in $\hbar\omega_{q}$ and represent $\dot{W}$ as a sum of two
different contributions: $\dot{W}=\dot{W}_{\mathrm{N}}+\dot{W}_{\mathrm{A}}$.
The normal energy loss rate $\dot{W}_{\mathrm{N}}$ is proportional to
$T_{e}-T$, which is a measure of deviation from the equilibrium,%
\begin{equation}
\dot{W}_{\mathrm{N}}=-\frac{\left(  T_{e}-T\right)  \hbar}{m_{e}A}%
\sum_{\mathbf{q}}\sum_{l,l^{\prime}}\bar{n}_{l^{\prime}}e^{-\Delta
_{l,l^{\prime}}/T_{e}}\tilde{\chi}_{l,l^{\prime}}^{\left(  r\right)
}S_{l,l^{\prime}}\left(  q,\omega_{l,l^{\prime}}\right)  , %
\label{e30}%
\end{equation}
Here
\[
\tilde{\chi}_{l,l^{\prime}}^{\left(  r\right)  }=\frac{m_{e}q}{2\rho\hbar
T_{e}}\left\vert \left(  U_{q}\right)  _{l,l^{\prime}}\right\vert ^{2}.
\]
This contribution originates from expansion of the exponential function in
$\hbar\omega_{q}\left(  1/T-1/T_{e}\right)  $.

It is conventional to represent the energy loss as $\dot{W}_{\mathrm{N}%
}=-\left(  T_{e}-T\right)  \tilde{\nu}_{\mathrm{N}}^{\left(  r\right)  }$,
where $\tilde{\nu}_{\mathrm{N}}^{\left(  r\right)  }$ is the energy relaxation
rate of an electron. Rearranging terms with $l<l^{\prime}$ ($\Delta
_{l,l^{\prime}}<0$), as described in the previous Section, one can find
\[
\tilde{\nu}_{\mathrm{N}}^{\left(  r\right)  }=\frac{\hbar}{m_{e}A}%
\sum_{\mathbf{q}}\sum_{l}\bar{n}_{l}\tilde{\chi}_{l,l}^{\left(  r\right)
}S_{l,l}\left(  q,0\right)  +
\]%
\begin{equation}
+\text{ }\frac{\hbar}{m_{e}A}\sum_{\mathbf{q}}\sum_{l>l^{\prime}}\left(
\bar{n}_{l}+\bar{n}_{l^{\prime}}e^{-\Delta_{l,l^{\prime}}/T_{e}}\right)
\tilde{\chi}_{l,l^{\prime}}^{\left(  r\right)  }S_{l,l^{\prime}}\left(
q,\omega_{l,l^{\prime}}\right) .  \label{e31}%
\end{equation}
The normal contribution $\tilde{\nu}_{\mathrm{N}}^{\left(  r\right)  }$ is
always positive, which means positive dissipation ($\dot{W}_{\mathrm{N}}<0$)
regardless of subband occupancies $\bar{n}_{l}$.

An anomalous contribution $\dot{W}_{\mathrm{A}}$ appears when expanding
$S_{l,l^{\prime}}\left(  q,\omega_{l,l^{\prime}}-\omega_{q}\right)  $ of
Eq.~(\ref{e29}) in $\omega_{q}$ and setting $e^{-\hbar\omega_{q}\left(
1/T-1/T_{e}\right)  }$ to unity. The rearrangement of terms with $l<l^{\prime
}$ ($\Delta_{l,l^{\prime}}<0$) based on the property of Eq.~(\ref{e21}) yields
\begin{eqnarray}
\dot{W}_{\mathrm{A}}=\frac{2TT_{e}}{m_{e}A}\sum_{l>l^{\prime}}\left(
\bar{n}_{l}-\bar{n}_{l^{\prime}}e^{-\Delta_{l,l^{\prime}}/T_{e}}\right)
\times \nonumber \\
\times \sum_{\mathbf{q}}\tilde{\chi}_{l,l^{\prime}}^{\left(  r\right)  }%
S_{l,l^{\prime}}^{\prime}\left(  q,\omega_{l,l^{\prime}}\right) .
\,\,\,\,\,\,\, \,\,\,\,\,\,\,\,\,\,\,\,\, \,\,\,\,\, \label{e32}%
\end{eqnarray}
Here $\bar{n}_{l}-\bar{n}_{l^{\prime}}e^{-\Delta_{l,l^{\prime}}/T_{e}}$
represents an additional measure of deviation from the equilibrium induced by
the MW. For equilibrium distribution of fractional occupancies $\bar
{n}_{l}$, the anomalous term equals zero, but for occupancies $\bar{n}_{2}>\bar{n}%
_{1}e^{-\Delta_{2,1}/T_{e}}$ induced by the MW resonance, $\dot{W}%
_{\mathrm{A}}$ can lead to negative energy dissipation of the electron system.
In Eq.~(\ref{e32}), one can use the approximate expression for $S_{l,l^{\prime
}}^{\prime}\left(  q,\omega_{l,l^{\prime}}\right)  $ given in Eq.~(\ref{e23}).
According to Eqs.~(\ref{e23}) and (\ref{e32}), the sign of $-\dot
{W}_{\mathrm{A}}$ coincides with the sign of $\omega_{2,1}-\left(  n^{\prime
}-n\right)  \omega_{c}$ which can be negative or positive depending on the
magnetic field. Since $\Gamma_{l,n;l^{\prime},n^{\prime}}\ll\hbar\omega
_{c}<\Delta_{2,1}$, the contribution $\dot{W}_{\mathrm{A}}$ is mostly
exponentially small with the exception of magnetic fields where $\Delta
_{l,l^{\prime}}-\left(  n^{\prime}-n\right)  \hbar\omega_{c}\lesssim
\Gamma_{l,n;l^{\prime},n^{\prime}}$.

The appearance of negative corrections to energy dissipation under the
condition $\bar{n}_{2}-\bar{n}_{1}e^{-\Delta_{2,1}/T_{e}}>0$ can be explained
quite easily. The negative anomalous contribution ($\dot{W}_{\mathrm{A}}>0$)
corresponds to $\left(  n^{\prime}-n\right)  \hbar\omega_{c}>\Delta_{2,1}$.
For narrow Landau levels, this means that scattering from the excited subband
($l=2$) to the ground subband ($l^{\prime}=1$) is accompanied by destruction of
a ripplon, while the corresponding scattering back from the ground subband to the excited subband
is accompanied by creation of a ripplon. When $\bar{n}_{2}=\bar{n}
_{1}e^{-\Delta_{2,1}/T_{e}}$, these two processes compensate each other in the
expression for $\dot{W}_{\mathrm{A}}$. If $\bar{n}_{2}>\bar{n}_{1}
e^{-\Delta_{2,1}/T_{e}}$, destruction of ripplons dominates, which leads to
negative dissipation. In the opposite case, when
$\left(  n^{\prime}-n\right)  \hbar\omega_{c}<\Delta_{2,1}$, creation of
ripplons dominates, which results in additional positive dissipation. It
should be noted that the negative contribution to energy dissipation and
negative momentum dissipation occur at the opposite sides of the
point\ $\hbar\omega_{c}=\Delta_{2,1}/m$. Comparing Eq.~(\ref{e17}) with
Eqs.~(\ref{e28}) and (\ref{e29}) one can conclude that the origin of this
difference is the negative sign of the Doppler-shift correction in the ripplon
excitation spectrum considered in the center-of-mass reference frame:
$E_{\mathbf{q}}^{\left(  r\right)  }=\hbar\omega_{\mathbf{q}}-\hbar
\mathbf{qV}_{\mathrm{av}}$.

Consider now the energy loss rate of SEs due to electron scattering at vapor
atoms. In this case, the interaction Hamiltonian is proportional to the
density fluctuation operator of vapor atoms $\rho_{\mathbf{K}}=\sum
_{\mathbf{K}^{\prime}}a_{\mathbf{K}^{\prime}-\mathbf{K}}^{\dag}a_{\mathbf{K}%
^{\prime}}$, where $\mathbf{K=}\left\{  K_{z},\mathbf{q}\right\}  $ represents
the momentum exchange between an electron and a scatterer. In terms of
$S_{l,l^{\prime}}^{(0)}\left(  q,\omega\right)  $, the energy loss rate
per an electron can be obtained as
\begin{eqnarray}
\dot{W}=-\frac{\left( V^{(a)}\right) ^{2}}{A^{2}L_{z}^{2}\hbar}\sum_{l,l^{\prime}}\bar{n}_{l}%
\sum_{\mathbf{K}^{\prime},\mathbf{K}}\varkappa_{\mathbf{K,K}^{\prime}%
}\left\vert \left(  e^{iK_{z}z_{e}}\right)  _{l,l^{\prime}}\right\vert
^{2} \times \nonumber \\
\times N_{\mathbf{K}^{\prime}}^{\left(  a\right)  }S_{l,l^{\prime}}\left(
q,\omega_{l,l^{\prime}}-\varkappa_{\mathbf{K,K}^{\prime}}\right)  ,
\,\,\,\,\,\,\,\,\,\,\,\,\,
\label{e33}%
\end{eqnarray}
where $\hbar\varkappa_{\mathbf{K,K}^{\prime}}=\varepsilon_{\mathbf{K}^{\prime
}-\mathbf{K}}^{\left(  a\right)  }-\varepsilon_{\mathbf{K}^{\prime}}^{\left(
a\right)  }$ is the energy exchange at a collision. In order to obtain
$\dot{W}_{\mathrm{N}}$ and $\dot{W}_{\mathrm{A}}$, we shall firstly rewrite
Eq.~(\ref{e33}) trivially as a sum of two
identical halves. Then, in the second half, the running indices $\mathbf{K}%
^{\prime},\mathbf{K}$ will be substituted as $\mathbf{K}^{\prime
}-\mathbf{K\rightarrow\tilde{K}}^{\prime}$, and $\mathbf{K\rightarrow
-\tilde{K}}$, which changes the sign of the energy exchange, $\varkappa
_{\mathbf{K,K}^{\prime}}\rightarrow-\varkappa_{\mathbf{\tilde{K},\tilde{K}%
}^{\prime}}$. The next steps are the same as those resulting in Eq.~(\ref{e29}%
). Interchanging the running indices $l\rightleftarrows l^{\prime}$ in the
second half, and using the basic property of $S_{l,l^{\prime}}^{(0)}\left(
q,\omega\right)  $ one can find
\begin{eqnarray}
\dot{W}=-\frac{\left( V^{(a)}\right) ^{2}}{2A^{2}L_{z}^{2}\hbar}\sum_{l,l^{\prime}}\sum
_{\mathbf{K}^{\prime},\mathbf{K}}\varkappa_{\mathbf{K,K}^{\prime}}\left\vert
\left(  e^{iK_{z}z_{e}}\right)  _{l,l^{\prime}}\right\vert ^{2}N_{\mathbf{K}%
^{\prime}}^{\left(  a\right)  } \nonumber \\
\times S_{l,l^{\prime}}\left(  q,\omega_{l,l^{\prime}}-\varkappa
_{\mathbf{K,K}^{\prime}}\right)  \,\,\,\,\,\,\,\,\,\,\,
\,\,\,\,\,\,\,\,\,\,\,\, \,\,\,\,\,\,\,\,\,\,\,\,\,\,\,\,\,\, \nonumber \\
\times \left[  \bar{n}_{l}-\bar{n}_{l^{\prime}%
}e^{-\Delta_{l,l^{\prime}}/T_{e}}e^{-\hbar\varkappa_{\mathbf{K,K}^{\prime}%
}\left(  1/T-1/T_{e}\right)  }\right] . \,\,\,\,\,\,\,\,\,\, \label{e34}%
\end{eqnarray}
This equation is more convenient for expansion in $\varkappa_{\mathbf{K,K}%
^{\prime}}$ than Eq.~(\ref{e33}).

Expanding Eq.~(\ref{e34}) in $\varkappa_{\mathbf{K,K}^{\prime}}$, one can find
again that $\dot{W}=\dot{W}_{\mathrm{N}}+\dot{W}_{\mathrm{A}}$, where $\dot
{W}_{\mathrm{N}}$ and $\dot{W}_{\mathrm{A}}$ have the same forms as that given
in Eqs.~(\ref{e30}) and (\ref{e32}), where $\tilde{\chi}_{l,l^{\prime}%
}^{\left(  r\right)  }$ should be substituted for
\begin{equation}
\text{\ }\tilde{\chi}_{l,l^{\prime}}^{\left(  a\right)  }=\frac{m_{e}\left( V^{(a)}%
\right) ^{2}}{2A L_{z}^{2}TT_{e}\hbar}\sum_{\mathbf{K}^{\prime},K_{z}}\left(
\varkappa_{\mathbf{K,K}^{\prime}}\right)  ^{2}N_{\mathbf{K}^{\prime}}^{\left(
a\right)  }\left\vert \left(  e^{iK_{z}z}\right)  _{l^{\prime},l}\right\vert
^{2}. \label{e35}%
\end{equation}
Using the condition $K^{\prime}\gg K$, this equation can be simplified as
\begin{equation}
\tilde{\chi}_{l,l^{\prime}}^{\left(  a\right)  }=\nu_{0}^{(a)}\frac{m_{e}}%
{M}\left(  u_{l,l^{\prime}}\frac{E_{R}}{T_{e}}+x_{q}\frac{\hbar\omega_{c}%
}{T_{e}}p_{l,l^{\prime}}\right)  , \label{e36}%
\end{equation}
where
\[
u_{l,l^{\prime}}=\frac{a_{B}^{2}B_{11}}{C_{l^{\prime}l}},\text{ \ \ }%
C_{l^{\prime}l}^{-1}=\frac{1}{L_{z}}\sum_{K_{z}}K_{z}^{2}\left\vert \left(
e^{iK_{z}z_{e}}\right)  _{l^{\prime}l}\right\vert ^{2}.
\]
Expressions for $C_{l^{\prime}l}^{-1}$ and $B_{l,l^{\prime}}^{-1}$ convenient
for numerical evaluations were given in Refs.~\onlinecite{SaiAok-78,MonKonKon-07}.

The energy loss $\dot{W}$ transferred to vapor atoms and ripplons is balanced
by the energy taken from the MW field: $\dot{W}=\left(  \bar{n}_{1}-\bar
{n}_{2}\right)  \Delta_{2,1}r$, where $r$ is the MW excitation rate defined by%
\begin{equation}
r=\frac{1}{2}\frac{\Omega_{R}^{2}\gamma}{\left(  \omega-\omega_{2,1}\right)
^{2}+\gamma^{2}}, \label{e37}%
\end{equation}
where $\gamma$ is the half-width of the MW resonance, and $\Omega_{R}$ is the Rabi
frequency proportional to the amplitude of the MW field. It is clear that
negative contribution of $\dot{W}_{\mathrm{A}}$ will be compensated by an
increase in $\dot{W}_{\mathrm{N}}$ due to electron heating.

Some useful expressions for the SE energy relaxation rate obtained for arbitrary
subband occupancies are given in the Appendix. It should be noted that
negative contributions to energy dissipation discussed above appear only for
quasi-elastic iter-subband scattering. For SEs above superfluid $^{4}%
\mathrm{He}$, there are inelastic inter-subband scattering processes
accompanying by simultaneous emission of two short wave-length
ripplons~\cite{MonKon-04,MonSokStu-10}. These processes
cause strong additional energy relaxation. Experiments of
Refs.~\onlinecite{KonKon-09,KonKon-10}, were performed for SEs on the free surface of
Fermi-liquid $^{3}\mathrm{He}$. For such a substrate, short wavelength
capillary waves with $q\gtrsim10^{7}$ are so heavily damped that even the
existence of ripplons with such wave-numbers is doubtful.

\section{Results and discussions}

\subsection{Vapor atom scattering regime}

Electron scattering at vapor atoms represents the most simple case for the
magnetotransport theory, because the collision broadening of Landau levels of the same
subband ($\Gamma_{l}$) is independent of the level number $n$. The same is
obviously valid for the broadening of the generalized factor $S_{l,l^{\prime}%
}\left(  q,\omega\right)  $, which now can be denoted as $\Gamma_{l;l}$.
Additionally, the parameter defined in Eq.~(\ref{e24}) has a very simple form
$\lambda_{l,n;l^{\prime},n+m}^{\left(  a\right)  }=\nu_{0}^{(a)}%
p_{l,l^{\prime}}\left(  2n+1+m\right)  $ which greatly simplifies evaluations.

Consider electron temperature as a function of the magnetic field. It is
defined by the energy balance equation which contains the MW excitation rate
$r$ given in Eq.~(\ref{e37}). In turn, $r$ depends on the half-width of the MW
resonance $\gamma$, which was studied theoretically with no magnetic field and
under a parallel magnetic field~\cite{And-78}. If $\mathbf{B}$ is
applied perpendicular to the surface, $\gamma$ should also have
1/B-oscillating terms, because inter-subband scattering increases when
$\Delta_{2,1}/\hbar\omega_{c}\rightarrow m$. In our numerical evaluation, we
shall use a qualitative extension of the result obtained for $B=0$.
According to this result, $\gamma$ contains the contribution
from intra-subband scattering $\gamma_{22-11}$ and the contribution from
inter-subband scattering $\gamma_{2,1}=\bar{\nu}_{2\rightarrow1}/2$. Under the
magnetic field applied normally, electron scattering is enhanced by the factor
$\hbar\omega_{c}/\sqrt{\pi}\Gamma_{l}$~\cite{AndUem-74}. Therefore, we can use an
approximation%
\[
\gamma_{22-11}\approx\frac{\nu_{0}^{(a)}\hbar\omega_{c}}{2\sqrt{\pi}%
\Gamma_{2,1}}\left[  p_{2,2}+p_{1,1}-2p_{2,1}\right]  ,
\]
where $\Gamma_{2,1}$ numerically is rather close to $\Gamma_{1}=\hbar
\sqrt{2\omega_{c}\nu_{0}^{(a)}/\pi}$.
As for the oscillatory part $\bar{\nu }_{2\rightarrow1}$
entering $\gamma_{2,1}$, we shall use the exact form of Eq.~(\ref{e14}).

\begin{figure}[tbp]
\begin{center}
\includegraphics[width=9.5cm]{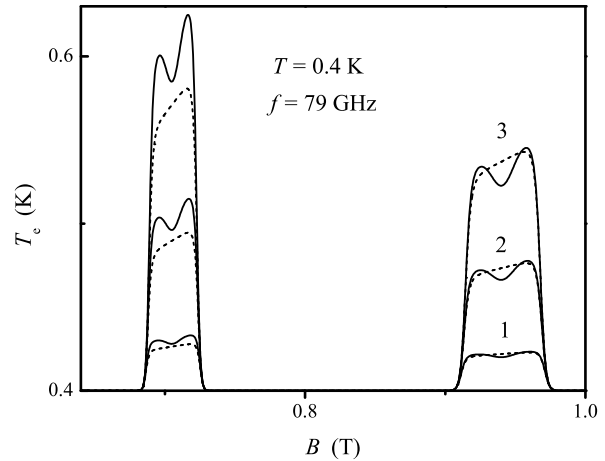}
\end{center}
\caption{Electron temperature vs the magnetic field for three levels of MW
irradiation estimated at $B=1\,\mathrm{T}$: $r=10^{5}\,\mathrm{s}^{-1}$ (1),
$3\cdot10^{5}\,\mathrm{s}^{-1}$ (2), and $5\cdot 10^{5}\,\mathrm{s}^{-1}$ (3).
The solid curve was calculated for the model of $\gamma (B)$ discussed in
the text, the dashed curve represents the case
$r(B)=\mathrm{const}$.} \label{f1}
\end{figure}

It should be noted that the oscillatory part of $\gamma$ is not large because
$p_{2,1}\simeq0.14$. Still, it leads to some important consequences for
electron temperature as a function of the magnetic field shown in Fig.~\ref{f1}.
Solid curves represent results of numerical evaluations for the two-subband
model taking into account oscillatory corrections to the MW resonance
half-width $\gamma$, as described above. In this case, the electron
temperature has small local minima at $\Delta_{2,1}/\hbar\omega_{c}\rightarrow
m$ due to oscillatory decrease in $r\propto1/\gamma$. Three typical values of
the Rabi frequency are chosen to provide MW excitation rate levels of
$10^{5}\,\mathrm{s}^{-1}$, $3\cdot10^{5}\,\mathrm{s}^{-1}$ and $5\cdot
10^{5}\,\mathrm{s}^{-1}$ at $B=1\,\mathrm{T}$. For a model with a constant MW
excitation rate $r$,\ which is applicable when inhomogeneous broadening
dominates, the corresponding results are shown by dashed curves. At
$\Delta_{2,1}/\hbar\omega_{c}\approx m$ these curves are nearly straight lines
(without minima). For both models, the shape of curves describing the
oscillatory increase of electron temperature has asymmetry with regard to the
point $\Delta_{2,1}/\hbar\omega_{c}=m$. This asymmetry is due to the negative
correction of the anomalous term $\dot{W}_{\mathrm{A}}$ leading to additional
heating of the electron system at $\hbar\omega_{c}>\Delta_{2,1}/m^{\ast}$.
The asymmetry increases strongly with the MW excitation rate $r$ and with
$m^{\ast}(B)$.

\begin{figure}[tbp]
\begin{center}
\includegraphics[width=9.5cm]{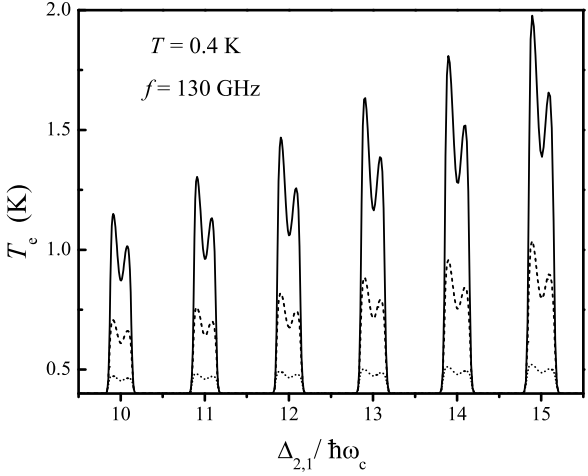}
\end{center}
\caption{Electron temperature vs $\Delta_{2,1}/\hbar\omega_{c}$
for the MW of higher resonant frequency, $f=130\,\mathrm{GHz}$.
Three levels of MW irradiation are the same as those described in the caption
of Fig.~\ref{f1}. The all curves were calculated for the model of $\gamma (B)$
discussed in the text. } \label{f2}
\end{figure}

Electron heating increases with $m^{\ast}$ (lowering $B$), and, for
the excitation rate $r=5\cdot10^{5}\,\mathrm{s}^{-1}$ at $B=1\,\mathrm{T}$,
the two-subband model fails at $m^{\ast}>11$. The applicability range of the
two-subband model can be extended by using a stronger holding electric field
which increases $\Delta_{2,1}$. In Fig.~\ref{f2}, electron temperature is shown
as a function of the parameter $\Delta_{2,1}/\hbar\omega_{c}\propto1/B$ for a
substantially higher MW frequency used in experiments on
SEs~\cite{KonIssKon-08}. For the solid curve, the two-subband model is applicable up
to $m^{\ast}=15$.

MW heating affects strongly the shape of conductivity oscillations because
$\nu_{\mathrm{eff}}$ depends on electron temperature. For example, the normal
contribution to the effective collision frequency can be represented as%
\begin{equation}
\nu_{\mathrm{N},intra}^{\left(  a\right)  }=\frac{\nu_{0}^{(a)}\hbar^{2}%
\omega_{c}^{2}}{2\sqrt{\pi}T_{e}}\sum_{l}\frac{\bar{n}_{l}p_{l,l}}%
{\Gamma_{l;l}}\coth\left(  \frac{\hbar\omega_{c}}{2T_{e}}\right)  , \label{e38}%
\end{equation}%
\begin{eqnarray}
\nu_{\mathrm{N},inter}^{\left(  a\right)  }=\frac{\nu_{0}^{(a)}\hbar^{2}%
\omega_{c}^{2}}{2\sqrt{\pi}T_{e}}\sum_{l>l^{\prime}}\left(  \bar{n}_{l}%
+\bar{n}_{l^{\prime}}e^{-\Delta_{l,l^{\prime}}/T_{e}}\right)  \frac
{p_{l,l^{\prime}}}{\Gamma_{l;l^{\prime}}}\times \nonumber \\
\times\left[  \coth\left(  \frac{\hbar\omega_{c}}{2T_{e}}\right)
F_{l,l^{\prime}}\left(  \omega_{c}\right)  +H_{l,l^{\prime}}\left(  \omega
_{c}\right)  \right]  ,\,\,\,\,\,\,\, \label{e39}%
\end{eqnarray}
where the new functions%
\begin{equation}
F_{l,l^{\prime}}\left(  \omega_{c}\right)  =\sum_{m=1}^{\infty}\exp\left[
-\frac{\hbar^{2}\left(  \omega_{l,l^{\prime}}-m\omega_{c}\right)  ^{2}}%
{\Gamma_{l;l^{\prime}}^{2}}\right]  ,\text{\ \ } \label{e40}%
\end{equation}%
\begin{equation}
H_{l,l^{\prime}}\left(  \omega_{c}\right)  =\sum_{m=1}^{\infty}m\exp\left[
-\frac{\hbar^{2}\left(  \omega_{l,l^{\prime}}-m\omega_{c}\right)  ^{2}}%
{\Gamma_{l;l^{\prime}}^{2}}\right]  \text{ } \label{e41}%
\end{equation}
defined for $l>l^{\prime}$ are independent of $T_{e}$. For narrow Landau
levels ($\Gamma_{l,l^{\prime}}\ll\hbar\omega_{c}$), the series defining
$F_{l,l^{\prime}}$ or $H_{l,l^{\prime}}$ can be approximated by a single term
with $m=m^{\ast}$, where $m^{\ast}$ depends on the magnetic field according to
the above noted rule: $m^{\ast}=\mathrm{round}\left(  \omega_{l,l^{\prime}%
}/\omega_{c}\right)  $.

The anomalous contribution to the effective collision frequency has a
different form%
\[
\nu_{\mathrm{A}}^{(a)}=-\frac{\nu_{0}^{(a)}\hbar^{2}\omega_{c}^{2}}{\pi^{1/2}%
}\sum_{l>l^{\prime}}\left(  \bar{n}_{l}-\bar{n}_{l^{\prime}}e^{-\Delta
_{l,l^{\prime}}/T_{e}}\right)  \frac{p_{l,l^{\prime}}}{\Gamma_{l;l^{\prime}%
}^{2}}\times
\]%
\begin{equation}
\times\left[  \coth\left(  \frac{\hbar\omega_{c}}{2T_{e}}\right)
\Phi_{l,l^{\prime}}\left(  \omega_{c}\right)  +\Theta_{l,l^{\prime}}\left(
\omega_{c}\right)  \right]  , \label{e42}%
\end{equation}
where functions $\Phi_{l,l^{\prime}}\left(  \omega_{c}\right)  $ and
$\Theta_{l,l^{\prime}}\left(  \omega_{c}\right)  $ are defined similar to
$F_{l,l^{\prime}}\left(  \omega_{c}\right)  $ and $H_{l,l^{\prime}}\left(
\omega_{c}\right)  $ of Eqs.~(\ref{e40}) and (\ref{e41}) respectively, with the
exception that their right sides contain the additional factor $2\hbar\left(
\omega_{l,l^{\prime}}-m\omega_{c}\right)  /\Gamma_{l;l^{\prime}}$ originated
from Eq.~(\ref{e23}). Similar equations for SE energy relaxation rate are given
in the Appendix.

\begin{figure}[tbp]
\begin{center}
\includegraphics[width=9.5cm]{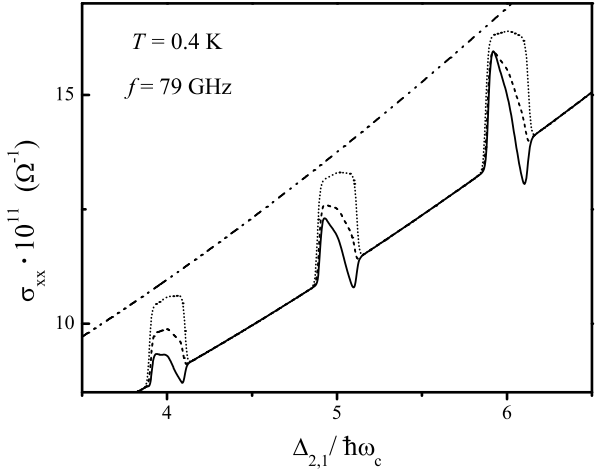}
\end{center}
\caption{Magnetoconductivity vs $\Delta_{2,1}/\hbar\omega_{c}\propto1/B$
for the MW of $f=79\,\mathrm{GHz}$. The dashed-dot-dot
line was calculated with no MW irradiation. Dotted, dashed and solid curves represent
three levels of MW irradiation ($r$) given in the caption of Fig.~\ref{f1}.
} \label{f3}
\end{figure}

Comparing $T_{e}$-dependencies of $\nu_{\mathrm{N}}^{\left(  a\right)  }$ and
$\nu_{\mathrm{A}}^{(a)}$ given in Eqs.~(\ref{e38}), (\ref{e39}) and (\ref{e42}),
we conclude that heating of the electron system reduces the normal
contribution to the effective collision frequency. In contrast with this, the
anomalous sign-changing correction $\nu_{\mathrm{A}}^{(a)}$, can be even
enhanced to a some extent with heating of SEs due to the factor $\coth\left(
\hbar\omega_{c}/2T_{e}\right)  $. Typical magnetoconductivity oscillations of
SEs calculated for the conditions of the experiment of Ref.~\onlinecite{KonKon-10}
are shown in Fig.~\ref{f3}. Electron temperature calculated for these curves was shown
in Fig.~\ref{f1} by solid curves. At low MW excitation ($r=10^{5}\,\mathrm{s}^{-1}$
at $B=1\,\mathrm{T}$), magnetooscillations of $\sigma_{xx}$ are just simple
maxima centered at $\Delta_{2,1}/\hbar\omega_{c}=m$, which agrees with
experimental observations. Between the commensurability conditions,
$\sigma_{xx}$ is suppressed, as compared to the dash-dot-dot line calculated
for zero MW power. This suppression is due to $\bar{n}_{2}\rightarrow$
$\bar{n}_{1}\rightarrow1/2$ and weaker scattering at the excited subband. The
increase in the decay rate $\bar{\nu}_{2\rightarrow1}$ at $\Delta_{2,1}%
/\hbar\omega_{c}\rightarrow m$ leads to a sharp decrease in $\bar{n}_{2}$,
which restores $\sigma_{xx}$ values obtained without the MW field. This
restoration is not complete if $T_{e}>T$, because $\nu_{\mathrm{N}%
,intra}^{\left(  a\right)  }$ decreases with heating, as discussed above. The
$\nu_{\mathrm{N},inter}^{\left(  a\right)  }$ is very small under these conditions.

\begin{figure}[tbp]
\begin{center}
\includegraphics[width=9.5cm]{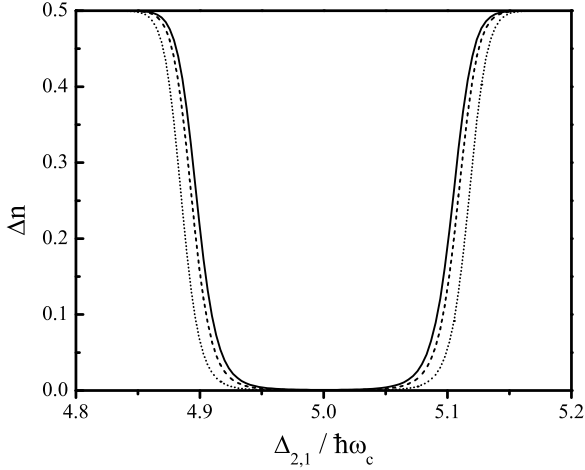}
\end{center}
\caption{Deviation of subband occupancies from the equilibrium distribution,
$\Delta n=\bar{n}_{2}-\bar{n}_{1}e^{-\Delta
_{2,1}/T_{e}}$, vs the parameter $\Delta_{2,1}/\hbar\omega_{c}\propto1/B$.
Dotted, dashed and solid curves represent
three levels of MW irradiation ($r$) given in the caption of Fig.~\ref{f1}.
} \label{f4}
\end{figure}

\begin{figure}[tbp]
\begin{center}
\includegraphics[width=9.5cm]{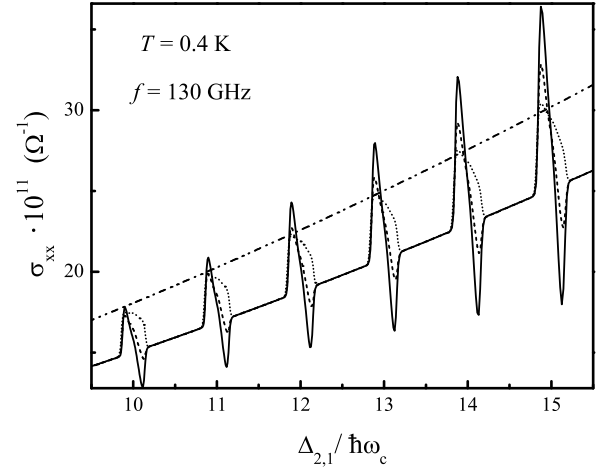}
\end{center}
\caption{Magnetoconductivity vs $\Delta_{2,1}/\hbar\omega_{c}$
for the MW field of higher resonant frequency, $f=130\,\mathrm{GHz}$. The dashed-dot-dot
line was calculated with no MW irradiation. Dotted, dashed and solid curves represent
three levels of MW irradiation ($r$) the same as those in Fig.~\ref{f2}.
} \label{f5}
\end{figure}

At higher MW excitation ($r=3\cdot10^{5}\,\mathrm{s}^{-1}$ and $5\cdot
10^{5}\,\mathrm{s}^{-1}$ at $B=1\,\mathrm{T}$), the shape of conductivity
oscillations is affected strongly by the anomalous term $\nu_{\mathrm{A}%
}^{(a)}$ leading to local minima at $\Delta_{2,1}/\hbar\omega_{c}>m$. The
$\nu_{\mathrm{A}}^{(a)}$ increases with $r$ because of two reasons. The first
reason is the increase in $\Delta n=\bar{n}_{2}-\bar{n}_{1}e^{-\Delta
_{2,1}/T_{e}}$ at higher MW excitation shown in Fig.~\ref{f4}. The second
reason is electron heating due to decay of the excited SE state which
increases $\coth\left(  \hbar\omega_{c}/2T_{e}\right)  $ of Eq.~(\ref{e42}).
Further shape evolution is shown in Fig.~\ref{f5} for lager $m^{\ast}$ and
$f=130\,\mathrm{GHz}$, where, according to Fig.~\ref{f2}, electron temperature
can take a value of about $2\,\mathrm{K}$. As expected, under these conditions
the anomalous contribution strongly increases. Near the commensurability
conditions $\Delta_{2,1}/\hbar\omega_{c}\rightarrow m$, the shape of
conductivity oscillations becomes similar to that observed for the
electron-ripplon scattering regime~\cite{KonKon-10} at $T=0.2\,\mathrm{K}$. It is
important that the results given in Fig.~\ref{f5} are still obtained in the
validity range of the two-subband model.

\begin{figure}[tbp]
\begin{center}
\includegraphics[width=9.5cm]{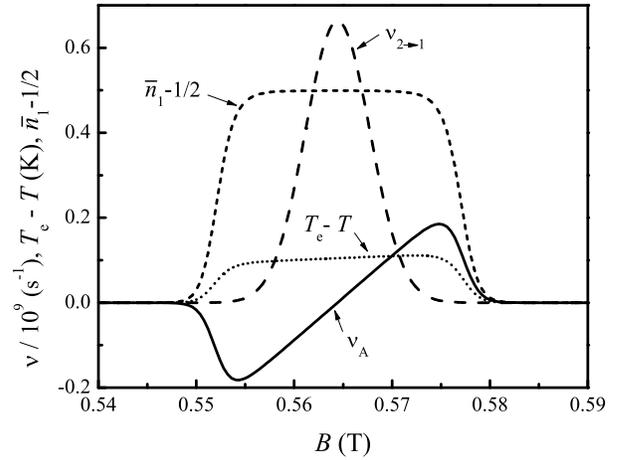}
\end{center}
\caption{Line shapes of $\bar{\nu}_{2\rightarrow1}$ (dashed), $ \nu_{\mathrm{A}}$ (solid),
$T_{e}-T$ (short-dotted), and $\bar{n}_{1}-1/2$ (short-dashed) as functions of $B$ near the
commensurability point with $m=5$, under the conditions: $T=0.4\,\mathrm{K}$, and
$r=3\cdot 10^{5}\, \mathrm{s^{-1}}$.
} \label{f6}
\end{figure}

It is instructive to compare the peak broadening of different quantities shown
in Fig.~\ref{f6}. The broadening of the decay rate $\bar{\nu}_{2\rightarrow1}$
coincides with $\Gamma_{2,1}$, which is an average of $\Gamma_{2}$ and
$\Gamma_{1}$. In contrast, such quantities as $\left\vert \nu_{\mathrm{A}%
}\right\vert $, $\bar{n}_{1}-1/2$, and $T_{e}-T$ rise in a much broader
magnetic field range having nearly the same width which does not represent the
broadening of Landau levels directly. We shall use this similarity in the
line widths of $T_{e}-T$ and $\bar{n}_{1}-1/2$ later, considering electron heating for
the electron-ripplon scattering regime.

\subsection{Electron-ripplon scattering regime}

For electron-ripplon scattering, the anomalous (sign-changing) contribution to
the effective collision frequency is induced by the MW resonance absolutely in
the same way, as it is for electron scattering at vapor atoms. In the case of
liquid $^{3}\mathrm{He}$, electron-ripplon scattering dominates at low
temperatures $T\leq0.2$, where the half-width of the MW resonance $\gamma$ is
substantially reduced. According to Eq.~(\ref{e37}), at the same amplitude of
the MW field, this decrease in $\gamma$ leads to a strong increase in the MW
excitation rate $r$ at the resonance $\omega=\omega_{2,1}$, which greatly magnifies
$\nu_{\mathrm{A}}$.

Unfortunately, the electron-ripplon scattering regime is much more difficult
for the analysis of the effect of electron heating than the vapor atom
scattering regime because of different reasons. First, the electron-ripplon
coupling $U_{q}$ has a very complicated form~\cite{MonKon-04}:%
\[
U_{q}\left(  z\right)  =\frac{\Lambda q}{z}\left[  \frac{1}{qz}-K_{1}\left(
qz\right)  \right]  +eE_{\bot}-\frac{\partial V_{e}^{(0)}}{\partial z_{e}},
\]
where $K_{1}\left(  x\right)  $ is the modified Bessel function of the second
kind, and $V_{e}^{(0)}\left(  z\right)  $ is the electron potential energy
over a flat surface. Therefore, it is impossible to obtain simple analytical
equations for the energy loss function $\dot{W}(T_{e})$. Moreover, if
$^{3}\mathrm{He}$ is used as the liquid substrate, there might be
contributions from other mechanisms of energy relaxation, which by now have no
strict theoretical descriptions.

\begin{figure}[tbp]
\begin{center}
\includegraphics[width=9.5cm]{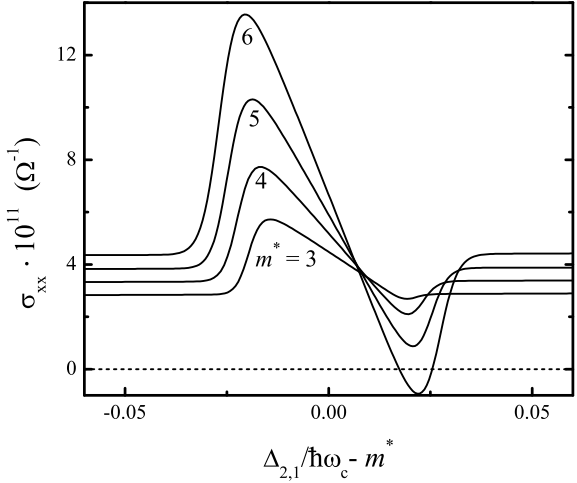}
\end{center}
\caption{Evolution of the $\sigma _{xx}(B)$ line shape near
commensurability points with the gradual increase in $m^{\ast}$ at $T=0.2\,\mathrm{K}$
and $f=79\,\mathrm{GHz}$.
} \label{f7}
\end{figure}

As indicated above, heating of SEs only increases the importance of the
anomalous contribution to the effective collision frequency. For
electron-ripplon scattering, it follows directly from Eqs.~(\ref{e25}%
)-(\ref{e27}). Therefore, in order to prove the possibility of existence of
zero resistance states due to non-equilibrium filling of the excited subband,
it is sufficient to show that negative $\sigma_{xx}$ can appear even without
electron heating. At $T_{e}=T=0.2\,\mathrm{K}$, the MW field amplitude, which
gave $r=10^{5}\,\mathrm{s}^{-1}$ (at $B=1\,\mathrm{T}$) for the vapor atom
scattering regime shown in Fig.~\ref{f3} (dotted curve), now gives
$r=2\cdot10^{6}\,\mathrm{s}^{-1}$, because the MW resonance line width
$2\gamma\simeq0.3\,\mathrm{GHz}$ due to inhomogeneous broadening~\cite{KonKon-10}.
This excitation rate is very high, because it leads to $\sigma_{xx}<0$ already
at $m^{\ast}=4$. For presentation of Fig.~\ref{f7}, we had chosen a two-times
lower excitation rate $r=10^{6}\,\mathrm{s}^{-1}$ independent of the magnetic
field. This figure shows the evolution of the line shape of conductivity
oscillations with the gradual increase in the integer parameter $m^{\ast}$. It
is quite convincing that even without heating of SEs the anomalous
contribution to the effective collision frequency increases strongly with
$m^{\ast}$, and the conductivity curve corresponding to $m^{\ast}=6$ enters
the negative conductivity regime in the vicinity of the minimum. This is in
accordance with experimental observations reported for the high
magnetic field range ($m^{\ast}<10$).

\begin{figure}[tbp]
\begin{center}
\includegraphics[width=9.5cm]{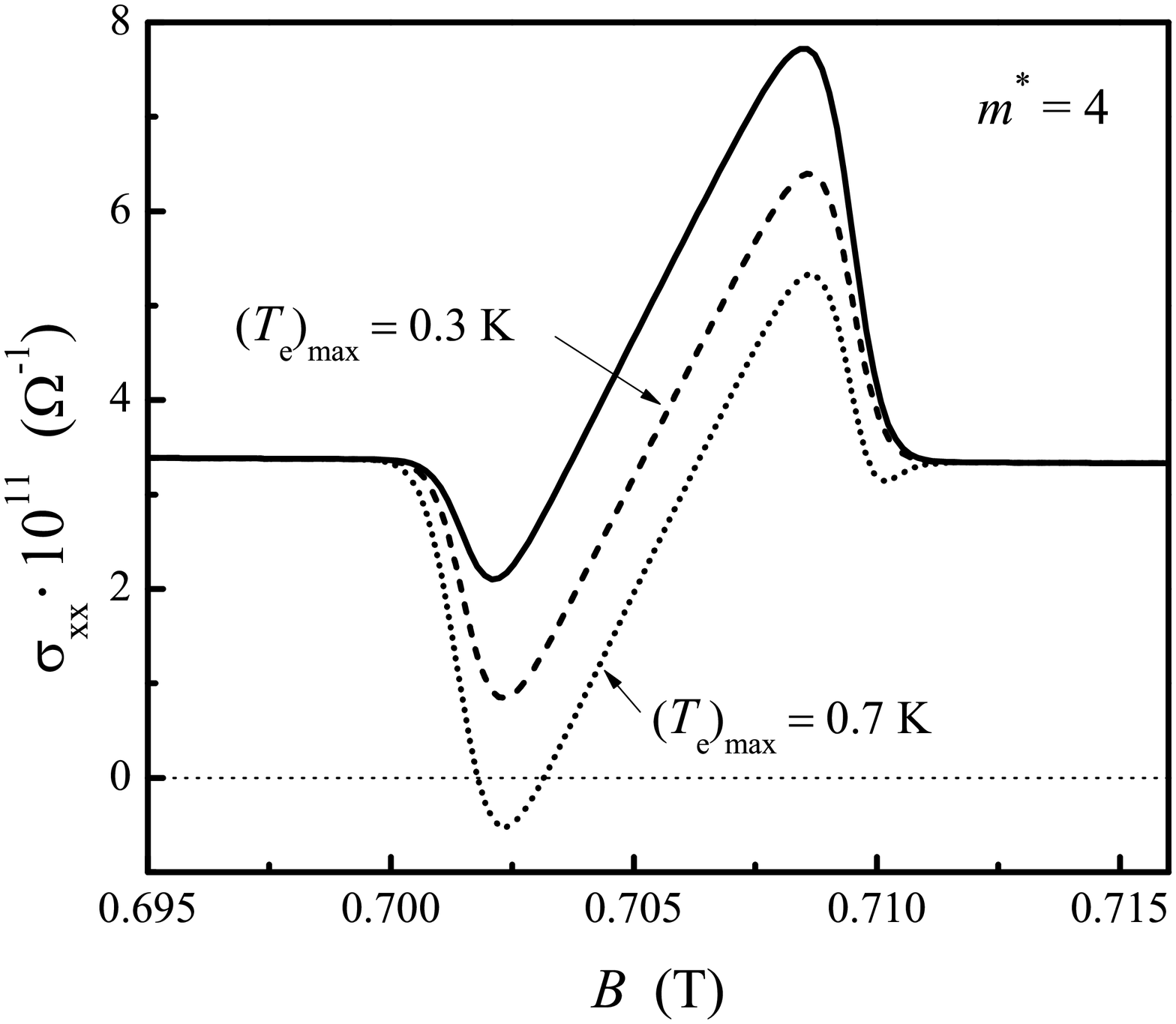}
\end{center}
\caption{Evolution of the $\sigma _{xx}(B)$ line shape near the
commensurability point $m^{\ast}=4$ with the increase in $\left(
\Delta T_{e}\right) _{\max}$
at $T=0.2\,\mathrm{K}$ and $f=79\,\mathrm{GHz}$: $\left(
\Delta T_{e}\right) _{\max}=0$ (solid), $0.1\,\mathrm{K}$ (dashed), and
$0.5\,\mathrm{K}$ (dotted).
} \label{f8}
\end{figure}

Maxima and minima of $\nu_{\mathrm{A}}(B)$ have the same amplitude. Without
heating of SEs, amplitudes of conductivity maxima obtained here are larger
than amplitudes of minima, because the normal contribution $\nu_{\mathrm{N}}$
increases at $\Delta_{2,1}/\hbar\omega_{c}\rightarrow m$ due to oscillations
of subband occupancies. Experimental curves~\cite{KonKon-10} show that at strong
MW power and large $m^{\ast}$ amplitudes of minima are larger.
This could be an indication of electron heating, because
$\nu_{\mathrm{N}}\sim1/T_{e}$. To analyze the effect of heating of SEs on
conductivity oscillations, we shall model electron temperature oscillations
using similarities in the line shapes of $T_{e}-T$ and $\bar{n}_{1}-1/2$ shown
in Fig.~\ref{f6}. In particular, we assume that an electron temperature peak is
described by $T_{e}\left(  B\right)  =T+2\left(  \Delta T_{e}\right)  _{\max
}\left[  \bar{n}_{1}(B,T)-1/2\right]  $, where the maximum elevation $\left(
\Delta T_{e}\right) _{\max}$ depends of $m^{\ast}$. We disregard the asymmetry of
the peak induced by $\dot{W}_{\mathrm{A}}$ because it does not lead to a
substantial change in final results. The results of such a model treatment of
the heating effect are shown in Fig.~\ref{f8}. They indicate that even moderate
heating of SEs affects strongly the shape of magnetooscillations, making
amplitudes of minima larger than amplitudes of maxima (dotted curve) in
accordance with experimental data.

In Fig.~\ref{f8}, we had chosen the excitation rate $r=10^{6}\,\mathrm{s}^{-1}$,
so that the initial curve (solid) calculated for $T_{e}=T$ have a small minima
with $\sigma_{xx}>0$. Then, we found that heating with $\left(  \Delta
T\right)  _{\max}=0.1\,\mathrm{K}$ strongly reduces conductivity extremes due to
$\nu_{\mathrm{N}}\sim1/T_{e}$, and moderate heating with $\left(  \Delta
T\right)  _{\max}=0.5\,\mathrm{K}$ leads to a minimum with $\sigma_{xx}<0$.
Therefore, decay heating of electrons, which occurs in the vicinity of the
commensurability conditions, helps to obtain zero resistance states. For
example, within the validity range of the two-subband model, electron
temperature peaks of about $2\,\mathrm{K}$ can reduce $\nu_{\mathrm{N}}$ by an
order of magnitude. Still, heating alone cannot make $\sigma_{xx}\leq0$. It is
only the anomalous contribution $\nu_{\mathrm{A}}$ which eventually leads to
negative conductivity and zero-resistance states. Without $\nu_{\mathrm{A}}$,
a conductivity dip would be an even function of the parameter $\omega
_{2,1}-m^{\ast}\omega_{c}$ with $\sigma_{xx}>0$. The existence of a
magnetoconductivity maxima at the opposite side of the point $\omega
_{2,1}-m^{\ast}\omega_{c}=0$ in experimental curves, which demonstrate
vanishing magnetoconductivity~\cite{KonKon-10}, is an additional evidence for a
sign-changing correction convincing that ZRS are realized
at the vanishing points.

It should be noted that at $T=0.2\,\mathrm{K}$, one-ripplon scattering processes
are not sufficient to prevent strong heating of the electron system at the
commensurability conditions. In particular, for $r=5\cdot10^{5}\,\mathrm{s}%
^{-1}$, estimation gives $\left(  T_{e}\right)  _{\max}\sim 3\,\mathrm{K}$ at
$m^{\ast}=4$. The model treatment of the heating effect discussed here allows
to draw conclusions about actual role of the electron heating in experiments
with SEs~\cite{KonKon-10}. For example, the firm conductivity maximum (without a minima)
observed for radiation power P of -25 dB at $m^{\ast}=4$ surely indicates that
electron heating is small or moderate under these conditions, and there is an
additional mechanism of energy relaxation at low ambient temperatures. We
speculate, that the magnetopolaronic effect and electron coupling with bulk
quasi-particles, giving a very small correction to the momentum relaxation rate
under experimental conditions, can contribute substantially to the energy
relaxation rate reducing electron temperature.

Experiments~\cite{KonKon-09,KonKon-10} are conducted for low surface electron
densities $n_{s}$ of about
$10^{6}\,\mathrm{cm}^{-2}$. Nevertheless, electron-electron interaction affects
noticeably experimental data. According to
Ref.~\onlinecite{DykKha-79}, under magnetic field an electron moves in a quasi-uniform
electric field of other electrons $E_{f}$ of fluctuational origin. Its average
value $E_{f}^{(0)}\simeq3\sqrt{T_{e}}n_{s}^{3/4}$ increases strongly with
electron temperature and density. The fluctuational electric field increases
the broadening of the DSF~\cite{MonTesWyd-02,MonKon-04} $\Gamma_{l,n;l^{\prime},n^{\prime}}\rightarrow
\sqrt{\Gamma_{l,n;l^{\prime},n^{\prime}}^{2}+x_{q}\Gamma_{C}^{2}}%
$, where $\Gamma_{C}=\sqrt{2}eE_{f}%
^{(0)}l_{B}\propto1/\sqrt{B}$. Thus, at $T_{e}=0.2\,\mathrm{K}$, and
$n_{s}=0.9\cdot10^{6}\,\mathrm{cm}^{-2}$, the Coulombic correction increases
$\Gamma_{l,n;l^{\prime},n^{\prime}}$ by about 1.3, if we assume $x_{q}\simeq
1$. If we take into account that the integrand of Eq.~(\ref{e24}) has a maximum
at $x_{q}\sim m^{\ast}+2$, the broadening increases approximately two times.
Therefore, a qualitative analysis indicates that the many-electron effect
becomes more important in the low magnetic field range where it increases the
width of conductivity oscillations and reduces amplitudes of maxima and
minima, which also agrees with experimental observations. Decay heating
increases the Coulombic correction to the broadening of magnetooscillations.
Still, a strict description of Coulombic effects on magnetoconductivity
oscillations requires a more careful study.

\section{Conclusion}

In summary, we have developed the theory of magnetoconductivity oscillations
in a multi-subband 2D electron system under MW irradiation of a resonant frequency.
We have shown that besides the quite obvious 1/B-modulation of conductivity, the
non-equilibrium filling of the excited subband induced by the MW resonance leads also
to sign-changing corrections to the effective collision frequency due to usual
inter-subband scattering. As the MW power goes up, the corresponding
increase in the amplitude of these sign-changing corrections can result
in the negative linear response conductivity and zero-resistance states.

Our theory is based on the self-consistent Born approximations, and it is
presented in a general way applicable for any quasi-elastic scattering
mechanism. As particular examples, we have considered two kinds of scatterers
which are typical for the electron system formed on the free surface of liquid
helium: helium vapor atoms and capillary wave quanta (ripplons). In the vapor
atom scattering regime, we found a strong 1/B-modulation of the electron
temperature, which increases sharply in the vicinity of commensurability
conditions. This decay heating is shown to enhance the effect of the
sign-changing terms in the longitudinal conductivity $\sigma_{xx}$.  
The evolution of the line-shape of conductivity oscillations with an increase of
the MW field amplitude is studied, taking into account heating of
surface electrons.

For the electron-ripplon scattering regime, we have shown that
magnetooscillations of large amplitude and the negative linear response
conductivity of SEs can easily appear under moderate MW excitation even for
cold SEs. The evolution of the line-shape of $\sigma_{xx}$ extremes caused by
an increase in the electron temperature is studied using a model treatment. We
believe that theoretical results presented in this work explain all major features of
MW-resonance-induced magnetooscillations observed in the system of
SEs on liquid helium, and support the suggestion~\cite{KonKon-09,KonKon-10}
that novel zero-resistance states are realized in such a system.

\appendix
\section{Energy relaxation rate}

Here we give final expressions for the energy relaxation rate of SEs due to scattering with vapor atoms. The normal $\tilde{\nu}^{(a)}_{\mathrm{N}}$ and anomalous
$\tilde{\nu}^{(a)}_{\mathrm{A}}$ energy relaxation rates are defined by the following
relationships: $\dot{W}_{\mathrm{N}}=-\left(  T_{e}-T\right)  \tilde{\nu
}^{(a)}_{\mathrm{N}}$, and $\dot{W}_{\mathrm{A}}=-T\tilde{\nu}^{(a)}_{\mathrm{A}}$. In
turn, $\tilde{\nu}^{(a)}_{\mathrm{N}}=\tilde{\nu}^{(a)}_{\mathrm{N},intra}+\tilde{\nu
}^{(a)}_{\mathrm{N},inter}$, where%
\begin{eqnarray}
\tilde{\nu}^{(a)}_{\mathrm{N},intra}=\frac{\nu_{0}^{(a)}m_{e}\hbar\omega_{c}E_{R}%
}{\pi^{1/2}MT_{e}}\sum_{l}\frac{\bar{n}_{l}}{\Gamma_{l;l}} \times \nonumber \\
\times \left[
u_{l,l}+\frac{\hbar\omega_{c}}{E_{R}}p_{l,l}\coth\left(  \frac{\hbar\omega
_{c}}{2T_{e}}\right)  \right]  , \label{e43}%
\end{eqnarray}%
\begin{widetext}
\begin{equation}
\tilde{\nu}^{(a)}_{\mathrm{N},inter}=\frac{\nu_{0}^{(a)}m_{e}\hbar\omega_{c}E_{R}%
}{\pi^{1/2}MT_{e}}\sum_{l>l^{\prime}}\frac{\bar{n}_{l}+\bar{n}_{l^{\prime}%
}e^{-\Delta_{l,l^{\prime}}/T_{e}}}{\Gamma_{l;l^{\prime}}}
\left\{  u_{l,l^{\prime}}F_{l,l^{\prime}}\left(  \omega_{c}\right)
+\frac{\hbar\omega_{c}}{E_{R}}p_{l,l^{\prime}}\left[  F_{l,l^{\prime}}\left(
\omega_{c}\right)  \coth\left(  \frac{\hbar\omega_{c}}{2T_{e}}\right)
+H_{l,l^{\prime}}\left(  \omega_{c}\right)  \right]  \right\}  ,
\label{e44}%
\end{equation}
\end{widetext}
functions $F_{l,l^{\prime}}$ and $H_{l,l^{\prime}}$ were given in
Eqs.~(\ref{e40}) and (\ref{e41}).

The anomalous energy relaxation rate can be represented as
\begin{widetext}
\begin{equation}
\tilde{\nu}^{(a)}_{\mathrm{A}}=\frac{2\nu_{0}^{(a)}m_{e}\hbar\omega_{c}E_{R}}%
{\pi^{1/2}M}\sum_{l>l^{\prime}}\frac{\left(  n_{l}-n_{l^{\prime}}%
e^{-\Delta_{l,l^{\prime}}/T_{e}}\right)  }{\Gamma_{l;l^{\prime}}^{2}%
}\left\{ u_{l,l^{\prime}}\Phi_{l,l^{\prime}}\left(  \omega_{c}\right)
+\frac{\hbar\omega_{c}}{E_{R}}p_{l,l^{\prime}}\left[  \Phi_{l,l^{\prime}%
}\left(  \omega_{c}\right)  \coth\left(  \frac{\hbar\omega_{c}}{2T_{e}%
}\right)  +\Theta_{l,l^{\prime}}\left(  \omega_{c}\right)  \right]
\right\}, \label{e45}%
\end{equation}
\end{widetext}
where $\Phi_{l,l^{\prime}}$ and $\Theta_{l,l^{\prime}}$ are the same as those
of Eq.~(\ref{e42}). For equilibrium subband occupancies, $\tilde{\nu
}_{\mathrm{A}}=0$. These equations were used for obtaining electron
temperature as a function of the magnetic field under the MW resonance.

\end{document}